\documentclass[a4paper,11pt]{article}
\usepackage{jheppub}
\usepackage{amsmath}
\usepackage{graphicx}
\usepackage{multirow}
\usepackage{bm}
\usepackage{bbm}
\usepackage{color}
\usepackage{slashed}
\usepackage{scalefnt}
\usepackage{grffile}

\allowdisplaybreaks
\title{Decay constants of $c \bar b$ mesons  involving the ten heavy flavor-changing currents at N$^3$LO QCD}

\author[a]{Wei Tao}
\author[a]{and Zhen-Jun Xiao\footnote{Corresponding author} }
\affiliation[a]{Department of Physics and Institute of Theoretical Physics, Nanjing Normal University, Nanjing 210023, China}
\emailAdd{taowei@njnu.edu.cn}
\emailAdd{xiaozhenjun@njnu.edu.cn}

\abstract{
Within the nonrelativistic QCD (NRQCD) framework, we complete  the three-loop calculations of the NRQCD renormalization constants and the matching coefficients,  for the heavy flavor-changing temporal vector, spatial-spatial tensor, spatial-temporal axial-tensor currents, which are coupled to the $P$-wave $c\bar b$ mesons.
We further study the ten decay constants for the $c\bar b$ mesons ($B_c,B_c^*,B_{c0}^*,B_{c1}$) coupled with the ten heavy flavor-changing currents involving (pseudo-)scalar, (axial-)vector and (axial-)tensor up to the next-to-next-to-next-to-leading order (N$^3$LO) of $\alpha_s$.
We obtain the six ratios of decay constants by approximating them to the corresponding ratios of matching coefficients.
We find the N$^3$LO QCD corrections to the six ratios of decay constants have good convergence and  weak scale-dependence.
We finally  predict the hierarchical
relationship among the ten decay constants for the $c\bar b$ mesons coupled with the ten  currents.	
}

\preprint{~}

\begin{document}

\maketitle
%\flushbottom	

\section{Introduction}

The family of beauty-charmed ($c\bar b$) mesons provides a unique laboratory for exploring the QCD dynamics of heavy quarks and  studying the strong and electroweak interactions~\cite{Gershtein:1994jw,Eichten:1994gt,Eichten:2019gig}, 
as it is the only meson system composed of two different heavy flavors in the
Standard Model (SM) of Particle Physics. 
Contrary to the charmonium ($c\bar c$) and bottomonium ($b\bar b$), 
the $c\bar b$ mesons cannot annihilate into gluons or photons.
The excited $c \bar b$ states, lying  below the threshold of the decay into the BD meson pair,   
will decay into the ground pseudo-scalar state $B_c(1S)$ via
electromagnetic radiative decays or hadronic transitions~\cite{Martin-Gonzalez:2022qwd}.
And the  ground pseudo-scalar state $B_c(1S)$  can only   decay weakly,  via the   processes  such as the transitions ${\bar b}\to{\bar c}W^{*+}$, ${ c}\to{ s}W^{*+}$, and the annihilation $c\bar b\to W^{*+}$ through a virtual $W$-boson~\cite{Eichten:2019gig,Gouz:2002kk}. 
As a result, the $c \bar b$ mesons are more stable, with narrower
widths than their counterparts in the $c\bar c$ and $b\bar b$ meson families~\cite{ParticleDataGroup:2022pth}.

The theoretical prediction for  beauty-charmed mesons was first made  about 40 years ago~\cite{Eichten:1980mw}. 
However, in the experimental aspect, the $c \bar b$ mesons are much less explored than
the charmonium and bottomonium due to the small production rate, as the dominant production mechanism requires 
the simultaneous production of 
  $c\bar c$ and $b\bar b$ quark pairs~\cite{Gershtein:1994jw,Chang:1992jb,LHCb:2014mvo,Chang:2003cr,Masetti:1995uk,CMS:2012oxa,Qian:2009zz}. 
The ground  pseudo-scalar state $B_c(1S)$ was first discovered  in 1998 by the CDF Collaboration~\cite{CDF:1998ihx} at the Fermilab Tevatron.
In 2014, the ATLAS Collaboration  reported    the observation of an excited $S$-wave $c\bar b$ meson state~\cite{ATLAS:2014lga},  which was interpreted as either  $B_c(2S)$ or $B_c^*(2S)$.
Five years later,  both   excited states $B_c(2S)$ and $B_c^*(2S)$ were independently observed by the CMS Collaboration~\cite{CMS:2019uhm} and  LHCb Collaboration~\cite{LHCb:2019bem},   
where the ground  vector state $B_c^*(1S)$ was also hinted but was  not directly observed    due to the difficulty in reconstructing  the low-energy photon emitted in the radiative transition $B_c^*(1S)\to B_c(1S)\gamma$.
%$P$-wave states in the $c\bar b$ meson family, such as the scalar meson $B_{c0}^*$ and the axial-vector meson $B_{c1}$, have not yet been reported in the experiment. However, they are expected to be discovered in the future~\cite{Eichten:2019gig}, thanks to the continuous improvement in experimental techniques for particle detection and identification.   
With the continuous improvement in experimental techniques for particle detection and identification, it is anticipated that more states in the $c\bar b$ meson family, such as the $P$-wave scalar meson $B_{c0}^*$ and  axial-vector meson $B_{c1}$, will  be discovered in   future experiments~\cite{Eichten:2019gig}. 
The progress of experiments demands the related theoretical research.
%the coupling of a current to a meson produced/annihilated via it
%the coupling of a meson to a current that produces/annihilates it
The decay constant, representing the coupling of a meson to a specific current that annihilates it~\cite{Becirevic:2003pn,Govaerts:1986ua}, plays an important role in understanding the meson decay properties.
The decay constant is not only a fundamental physical quantity describing the leptonic  
decay of a meson state, but is also an important input parameter related to distribution amplitudes, form factors, decay widths and
branching ratios for various decay channels~\cite{Li:2009tx,Yang:2021crs,Calibbi:2022ddo,Hazard:2016fnc,Grinstein:2015aua,Ball:1997rj,Ball:1998kk,Becirevic:2003pn,Yang:2007zt,Chai:2021pyp,Herdoiza:2006qv,Chen:2020qma,Aloni:2017eny,Braun:2003jg,Braun:2016wnx,Choi:2007yu,Verma:2011yw,Ball:1996tb,Ball:2004rg,Hatanaka:2008xj,Yang:2005gk,Cheng:2007mx,Yang:2008xw,Cheng:2007st,Cheng:2005nb,Cheng:2008gxa,Bayar:2009tux,Cheng:2013fba,Wang:2008da,Lu:2006fr,Sungu:2010zz,Han:2023pgf,Pullin:2021ebn,Wang:2015mxa,Wang:2012kw,Wang:2013ywc,Tao:2022qxa,Tao:2023mtw,Dominguez:1988wa}.
Moreover, the decay constant can aid in  extracting  the
CKM matrix element in the SM  and studying new physics beyond it~\cite{Hatton:2020vzp,Gracey:2022vqr,Blake:2016olu,Chizhov:2003qy,Chimirri:2023ovl,Bali:2017pdv,Becirevic:2013bsa,Jansen:2009yh,Jansen:2009hr,Domokos:2011dn,RBC-UKQCD:2008mhs,Godfrey:2015vda,Ball:2006eu}.

Decay constants for various mesons involving various currents have been studied  through various methods.
With  the lattice QCD, 
the  decay constants of the $P$-wave light and heavy-light mesons  coupled with the scalar, vector, axial-vector and axial-tensor currents can be found in refs.~\cite{Herdoiza:2006qv,McNeile:2006nv,Bali:2017pdv},
and the  decay constants for the $S$-wave  and $P$-wave  light mesons and heavy quarkonia   coupled with the tensor current are known from the literature~\cite{Becirevic:2013bsa,Jansen:2009yh,Jansen:2009hr,Chen:2020qma,Aloni:2017eny,Domokos:2011dn,Hatton:2020vzp,Hazard:2016fnc,Becirevic:2003pn,RBC-UKQCD:2008mhs,Glozman:2011gf,Braun:2003jg,Becirevic:1998jp,Braun:2016wnx}.
By the Bethe-Salpeter equation  (BSE) method,
the $P$-wave $B_{c0}^*$ and $B_{c1}$  decay constants involving the scalar and axial-vector currents were researched   in refs.~\cite{Wang:2007av,Guleria:2020kuy,Li:2018eqc}.
Using the quark model, 
the $S$-wave and $P$-wave light meson decay constants involving the axial-vector, tensor and  axial-tensor currents were calculated in refs.~\cite{Chizhov:2003qy,Amarante:1973xs,Choi:2007yu,Arifi:2022qnd,Cappiello:2010tu,Alvares:2011wb,Colangelo:2011xk},
the $P$-wave   heavy-light meson decay constants  involving the scalar, vector and axial-vector currents can be found in ref.~\cite{Cheng:2003sm},
the $P$-wave heavy quarkonium decay constants involving the scalar current were  computed in ref.~\cite{Godfrey:2015vda}, 
and the $P$-wave  $B_{c0}^*$ and $B_{c1}$   decay constants  involving the scalar, vector and axial-vector currents were investigated in refs.~\cite{Verma:2011yw,Chang:2018zjq,Wang:2008as}.
Within the  QCD sum rules, 
the  $S$-wave and $P$-wave  light  meson decay constants involving the scalar, vector, axial-vector, tensor and axial-tensor currents were calculated in the literature~\cite{Yang:2007zt,Ball:1997rj,Ball:1998kk,Bakulev:1999gf,Belyaev:1996fd,Broniowski:1998ws,Craigie:1981jx,Ball:1996tb,Ball:2004rg,Ball:2006eu,Chernyak:1983ej,Hatanaka:2008xj,Yang:2005gk,Cheng:2007mx,Yang:2008xw,Cheng:2007st,Cheng:2005nb,Cheng:2008gxa,Bayar:2009tux,Sundu:2011vz,Cheng:2013fba,Wang:2008da,Lu:2006fr,Maltman:1999jn,Cata:2008zc,Chai:2021pyp,Li:2009tx}, 
the $S$-wave and $P$-wave heavy-light meson decay constants  involving the scalar, axial-vector, tensor and axial-tensor currents can be found in refs.~\cite{Sungu:2010zz,Han:2023pgf,Pullin:2021ebn,Wang:2015mxa},
the   $P$-wave heavy quarkonium decay constants coupled with the tensor and axial-tensor currents were studied in the literature~\cite{Wang:2012gj}, 
and the $P$-wave $B_{c0}^*$ and  $B_{c1}$  decay constants coupled with the scalar and axial-vector currents can be obtained from refs.~\cite{Narison:2020wql,Wang:2012kw,Narison:2015nxh,Aliev:1992vp,Yazici:2016foi}.
Using the Heavy Quark Effective Theory (HQET),   
the $S$-wave heavy-light meson decay constants coupled with the tensor current were computed in refs.~\cite{Broadhurst:1994se,Bekavac:2009zc,Grinstein:2015aua,Campanario:2003ix,Grozin:2010ai}.
Based on the nonrelativistic (NRQCD) effective field theory, 
the $S$-wave and $P$-wave heavy quarkonium decay constants involving the scalar, vector,   axial-vector, tensor and axial-tensor currents  have been calculated in refs.~\cite{Chung:2020zqc,Chung:2021efj,Wang:2013ywc,Kiselev:2001xa},
and the $S$-wave $B_c$ and $B_c^*$ meson decay constants involving the heavy flavor-changing  pseudo-scalar, vector, axial-vector, tensor and axial-tensor currents were studied in refs.~\cite{Feng:2022ruy,Sang:2022tnh,Braaten:1995ej,Hwang:1999fc,Lee:2010ts,Onishchenko:2003ui,Chen:2015csa,Tao:2022qxa,Tao:2023mtw,Tao:2023vvf}.

The $c\bar b$ mesons are intermediate between $c\bar c$ and $b\bar b$ states both in mass and size,
so many features of the beauty-charmed  meson family can be inferred from what we know of the charmonium and
bottomonium systems~\cite{CMS:2019uhm,Gershtein:1994jw,Eichten:1994gt,Eichten:2019gig,Martin-Gonzalez:2022qwd}. 
Specifically, the $c\bar b$ meson family
shares the nonrelativistic dynamical property with both the $c\bar c$ and $b\bar b$ systems~\cite{LHCb:2019bem,Gershtein:1994jw,Ortega:2020uvc,Eichten:2019gig,Martin-Gonzalez:2022qwd}.
Thus, it is appropriate to study the  decay constants for the low-lying $c\bar b$ mesons using
the NRQCD effective field theory.

With the framework of NRQCD factorization, at the lowest order     in quark relative
velocity expansion, the meson decay constant in QCD can be factorized into the short-distance coefficient (the matching coefficient) multiplied
with the NRQCD long-distance matrix element (the wave function at the origin)~\cite{Bodwin:1994jh}.
The matching coefficients for various heavy flavor-changing currents involving two different quark masses have been calculated at various perturbative orders in the strong coupling constant $\alpha_s$.
The  one-loop matching coefficient of the heavy flavor-changing temporal axial-vector current was first calculated in ref.~\cite{Braaten:1995ej}.
The one-loop matching coefficients of  the heavy flavor-changing spatial vector and temporal axial-vector currents  allowing for higher-order relativistic corrections can be found in refs.~\cite{Hwang:1999fc,Lee:2010ts}.
Two-loop corrections to  the matching coefficients for the heavy flavor-changing  pseudo-scalar, spatial vector and temporal axial-vector currents are available in the literature~\cite{Onishchenko:2003ui,Chen:2015csa,Tao:2022qxa}.
At the N$^3$LO of $\alpha_s$, the matching coefficients have been numerically computed for the seven heavy flavor-changing currents: scalar~\cite{Tao:2022hos}, pseudo-scalar~\cite{Tao:2022hos,Tao:2023mtw}, spatial vector~\cite{Sang:2022tnh},  temporal axial-vector~\cite{Feng:2022ruy}, spatial axial-vector~\cite{Tao:2023mtw}, spatial-temporal tensor~\cite{Tao:2023vvf} and spatial-spatial axial-tensor~\cite{Tao:2023vvf}.

In this work, we will complete the three-loop matching between QCD and NRQCD for the remaining three heavy flavor-changing currents: temporal vector, spatial-spatial tensor and spatial-temporal axial-tensor, which are coupled to the $P$-wave $c\bar b$ mesons.
Furthermore, we will investigate the decay constants, for the low-lying $S$-wave pseudo-scalar meson $B_c(1^{1}S_0)$, vector meson $B_c^*(1^{3}S_1)$ and the low-lying $P$-wave scalar meson $B_{c0}^*(1^{3}P_0)$, axial-vector meson $B_{c1}(1^{3}P_1)$~\cite{Martin-Gonzalez:2022qwd}, coupled with the  scalar, pseudo-scalar,  vector, axial-vector, tensor and axial-tensor currents, up to N$^3$LO within the NRQCD factorization formalism.
The obtained N$^3$LO results concerning the matching coefficients and decay constants will test the perturbative convergence of the NRQCD effective theory. 
As  the $c\bar b$ mesons are less explored, our predictions for their decay constants offer a significant opportunity for a  thorough and comprehensive  study of their decay properties. 
Since only  the $S$-wave $c\bar b$ mesons  are known experimentally, these predictions will provide valuable information for the experimental discovery of more states within the $c\bar b$ meson family, particularly the $P$-wave $c\bar b$ states.

The layout of this paper is as follows.
In Sec.~\ref{Matchingformula}, we define the decay constants in QCD and introduce the  matching formula between QCD and NRQCD.
In Sec.~\ref{QCDvertexfunction},  we describe  our calculation procedure for  QCD vertex functions.
In Sec.~\ref{ZjNRQCD}, we  present the three-loop analytic results of   NRQCD current  renormalization constants and corresponding anomalous dimensions.
In Sec.~\ref{Matchingcoefficient}, we present the three-loop  numerical results of   matching coefficients.
In Sec.~\ref{Decayconstantratio}, we study N$^3$LO QCD corrections to 
the ratios of decay constants for $c\bar b$ mesons coupled with various currents.
We summarize in Sec.~\ref{Summary}.

%the  decay constant ratios of $c\bar b$ mesons involving various currents. 
%the ratios of $c\bar b$ decay constants involving various currents.

\section{Matching formula ~\label{Matchingformula}}

We consider  the decay constant $f_{c\bar b}^J$ of the meson $c\bar b$  coupled with the heavy  flavor-changing  current $J$, where $c\bar b$ can be the $S$-wave pseudo-scalar meson $B_c(0^-)(^1S_0)$, vector meson $B^*_c(1^-)(^3{S_1})$, and the $P$-wave scalar meson $B^*_{c0}(0^+)({^3P_0})$, axial-vector meson $B_{c1}(1^+)(^3P_1)$. 
And $J$ belongs to the ten currents:  scalar $(s)$, pseudo-scalar $(p)$, temporal vector $(v,0)$, spatial vector $(v,i)$, temporal axial-vector $(a,0)$, spatial axial-vector $(a,i)$, spatial-temporal tensor $(t,i0)$, spatial-spatial tensor $(t,ij)$, spatial-temporal axial-tensor $(t5,i0)$, spatial-spatial axial-tensor $(t5,ij)$, i.e.
\begin{align}\label{j10}
J\in\{s,p,(v,0),(v,i),(a,0),(a,i),(t,i0),(t,ij),(t5,i0),(t5,ij)\}.
\end{align}
The decay constant $f_{c\bar b}^J$ can be defined from the matrix element of the QCD current that annihilates the meson~\cite{Chetyrkin:2021qvd,Neubert:1992fk,Sun:2022hyk,Soni:2017wvy,Colquhoun:2015oha,Dowdall:2012ab,Wang:2022cxy,Broadhurst:1994se,Koenigstein:2016tjw,Wang:2012kw,Burakovsky:1997ci,Sundu:2011vz,Abreu:2020ttf,Lu:2006fr,Amarante:1973xs,Pullin:2021ebn,Chen:2020qma,Penin:2001ux,Hwang:2012nw,Chang:2020wvs,Chang:2018aut,Sungu:2010zz,Chung:2020zqc,Chung:2021efj,Narison:2015nxh,Hwang:1997ie,Wang:2015mxa,Leljak:2021pho,Wang:2012gj,Aloni:2017eny,Braguta:2008qe,Olpak:2016enb,Choi:2007yu,Cheng:2003sm,Chernyak:1983ej,Mikhailov:2020tta,Braun:2016wnx,Wang:2013ywc,Chizhov:2003qy,Cata:2009zza,Cata:2008zc,Domokos:2011dn,Colangelo:2011xk,Gorsky:2020bwi,Martin-Gonzalez:2022qwd}:
\begin{align}\label{decayconstants}
 \langle 0 |j_s |{B}^*_{c0}(q) \rangle &\doteq %{\rm i}
 f^{s}_{{B}^*_{c0}}m_{{B}^*_{c0}},  
 \nonumber \\
  \langle 0 |j_p |B_c(q) \rangle &\doteq %{\rm i}
  f^{p}_{B_c}m_{B_c}, 
   \nonumber \\
   \langle 0 |j_v^{\mu} |{B}^*_{c0}(q) \rangle &\doteq %{\rm i} 
   f^{v,0}_{{B}^*_{c0}}q^\mu, 
    \nonumber \\
 \langle 0 |j_v^{\mu} |B^*_c(q,\varepsilon) \rangle &\doteq f^{v,i}_{B_c^*}m_{B_c^*} \varepsilon^\mu,  
 \nonumber \\
\langle 0 |j_a^{\mu}|B_c(q)\rangle &\doteq %{\rm i}
f^{a,0}_{B_c}q^{\mu},
\nonumber \\
\langle 0|j_a^{\mu}|{B}_{c1}(q,\varepsilon)\rangle &\doteq f^{a,i}_{{B}_{c1}}m_{{B}_{c1}} \varepsilon^\mu, 
\nonumber  \\
  \langle 0 |j_t^{\mu\nu} |B^*_c(q,\varepsilon) \rangle &\doteq f^{t,i0} _{B_c^*}(q^\mu \varepsilon^\nu-q^\nu \varepsilon^\mu), 
  \nonumber  \\
 \langle 0 |j_t^{\mu\nu} |{B}_{c1}(q,\varepsilon) \rangle &\doteq f^{t,ij} _{{B}_{c1}}\epsilon^{\mu\nu\alpha\beta}q_\alpha \varepsilon_\beta,  
 \nonumber \\         
 \langle 0 |j_{t5}^{\mu\nu} |{B}_{c1}(q,\varepsilon) \rangle &\doteq  f^{t5,i0}_{{B}_{c1}}(q^\mu \varepsilon^\nu-q^\nu \varepsilon^\mu),   
 \nonumber \\
  \langle 0 |j_{t5}^{\mu\nu} |B^*_c(q,\varepsilon) \rangle &\doteq  f^{t5,ij}_{B_c^*}\epsilon^{\mu\nu\alpha\beta}q_\alpha \varepsilon_\beta,      
\end{align}
where $q$ refers to the total momentum of a $c\bar b$ meson, and $\varepsilon$ denotes the polarization vector of $B_c^*$ ($B_{c1}$). The QCD heavy flavor-changing currents involving the bottom quark $\bar b$ and charm quark $c$ read:
\begin{align}\label{QCDcurrents}
j_s     &= \bar{\psi}_b \psi_c,
\nonumber \\
j_p     &= \bar{\psi}_b \gamma_5 \psi_c,
\nonumber \\
j_v^\mu &= \bar{\psi}_b \gamma^\mu \psi_c,
\nonumber \\
j_a^\mu &= \bar{\psi}_b \gamma^\mu\gamma_5 \psi_c,
\nonumber \\
j_t^{\mu\nu} &= \bar{\psi}_b \sigma^{\mu\nu}\psi_c,
\nonumber \\
j_{t5}^{\mu\nu} &= \bar{\psi}_b  \sigma^{\mu\nu}\gamma_5 \psi_c,
\end{align}
where
$\sigma_{\mu\nu}=\frac{\rm i}{2}(\gamma_{\mu}\gamma_{\nu}-\gamma_{\nu}\gamma_{\mu})$.
The   QCD current components coupled to the decay constants can be expanded with respect to the small half relative spatial momentum $|\vec{k}|$ between the bottom and charm quarks~\cite{Hwang:1999fc,Piclum:2007an} as follows:
\begin{align}\label{expandcurrents}
j_s     &= \mathcal{C}_{s}\tilde{j}_s + {\mathcal O}(|\vec{k}|^{3}),
\nonumber \\
j_p     &= \mathcal{C}_{p}\tilde{j}_p + {\mathcal O}(|\vec{k}|^{2}),
\nonumber \\
j_v^0 &=\mathcal{C}_{v,0} \tilde{j}_v^0+ {\mathcal O}(|\vec{k}|^{3}),
\nonumber \\
j_v^i &= \mathcal{C}_{v,i}\tilde{j}_v^i + {\mathcal O}(|\vec{k}|^{2}),
\nonumber \\
j_a^0 &= \mathcal{C}_{a,0}\tilde{j}_a^0 + {\mathcal O}(|\vec{k}|^{2}),
\nonumber \\
j_a^i &= \mathcal{C}_{a,i}\tilde{j}_a^i + {\mathcal O}(|\vec{k}|^{3}),
\nonumber \\
j_t^{i0}    &= \mathcal{C}_{t,i0}\tilde{j}_t^{i0} + {\mathcal O}(|\vec{k}|^{2}),
\nonumber \\
j_t^{ij}    &= \mathcal{C}_{t,ij}\tilde{j}_t^{ij} + {\mathcal O}(|\vec{k}|^{3}),
\nonumber \\
j_{t5}^{i0}    &= \mathcal{C}_{t5,i0}\tilde{j}_{t5}^{i0} + {\mathcal O}(|\vec{k}|^{3}),
\nonumber \\
j_{t5}^{ij}    &= \mathcal{C}_{t5,ij}\tilde{j}_{t5}^{ij} + {\mathcal O}(|\vec{k}|^{2}),
\end{align}
with
\begin{align}\label{NRQCDcurrents}
 \tilde{j}_s     &= -\varphi_b^\dagger  \left(\frac{1}{2m_c} +\frac{1}{2m_b}\right)  \vec{k}\cdot\vec{\sigma} \chi_c,
 \nonumber \\
 \tilde{j}_p     &= -\varphi_b^\dagger \chi_c,
 \nonumber \\
\tilde{j}_v^0 &= \varphi_b^\dagger \left(\frac{1}{2m_c} -\frac{1}{2m_b}\right)\vec{k}\cdot\vec{\sigma} \chi_c,
\nonumber \\
\tilde{j}_v^i &= \varphi_b^\dagger \sigma^i \chi_c,
\nonumber \\
\tilde{j}_a^0 &= \varphi_b^\dagger \chi_c,
\nonumber \\
\tilde{j}_a^i &=\varphi_b^\dagger\left(\frac{\sigma^i \vec{k}\cdot\vec{\sigma}}{2m_c} -\frac{ \vec{k}\cdot\vec{\sigma} \sigma^i}{2m_b}\right) \chi_c  ,
\nonumber \\
\tilde{j}_{t}^{i0} &=\varphi_b^\dagger{\rm i}   \sigma^i \chi_c,
\nonumber \\
\tilde{j}_{t}^{ij}&= \varphi_b^\dagger{\rm i}\left(\frac{\sigma^i \sigma^j\vec{k}\cdot\vec{\sigma}}{2m_c}+\frac{\vec{k}\cdot\vec{\sigma} \sigma^i\sigma^j}{2m_b}  \right)\chi_c,
\nonumber \\
\tilde{j}_{t5}^{i0}&=\varphi_b^\dagger{\rm i}\left(\frac{\sigma^i \vec{k}\cdot\vec{\sigma}}{2m_c}+\frac{\vec{k}\cdot\vec{\sigma} \sigma^i}{2m_b}  \right)\chi_c,
\nonumber \\
\tilde{j}_{t5}^{ij} &=\varphi_b^\dagger{\rm i}   \sigma^i \sigma^j\chi_c,
\end{align}
denoting the corresponding components of NRQCD currents  at the lowest order in $|\vec{k}|$.
$\mathcal{C}_{J}$ is the matching coefficient between QCD and NRQCD for the heavy flavor-changing current $J$. %$J\in\{s,p,(v,0),(v,i),(a,0),(a,i),(t,i0),(t,ij),(t5,i0),(t5,ij)\}$. 
$\varphi_b^\dagger$ and $\chi_c$ represent 2-component Pauli spinor  fields annihilating the $\bar{b}$ and $c$ quarks, respectively.

With both the QCD and NRQCD currents in eq.~\eqref{expandcurrents} sandwiched between the vacuum state  and the free $c\bar{b}$ pair of on-shell heavy charm and bottom quarks with small relative velocity~\cite{Beneke:1997jm,Onishchenko:2003ui}, the matching formula for $J$ in eq.~\eqref{j10} can be written as
\begin{align} \label{matchingformula}
\sqrt{Z_{2,b}^\mathrm{OS} Z_{2,c}^\mathrm{OS} } \,Z_{J}^{\mathrm{OS}}\left(Z_{J}^{\overline{\mathrm{MS}}}\right)  \, \Gamma_{J} &=
\mathcal{C}_{J}\left(\overline{\mathcal{C}}_{J}\right) \, \sqrt{\widetilde{Z}_{2,b}^\mathrm{OS} \widetilde{Z}_{2,c}^\mathrm{OS} } \,
{\widetilde Z}_{J}^{-1} \, \widetilde{\Gamma}_{J}  + \cdots.
\end{align}
The ellipses  denote higher-order relativistic corrections in $|\vec{k}|$, which are not considered in this paper.
Furthermore, we have removed the soft, potential and ultrasoft contributions from both the left-hand side (lhs) and  right-hand side (rhs) of the matching formula, so that $\Gamma_{J}$ denotes the unrenormalized  on-shell   vertex function with current $J$ in the hard integration region of QCD~\cite{Beneke:1997jm,Onishchenko:2003ui} while  $\widetilde{\Gamma}_{J}$ becomes the tree-level on-shell  vertex function with current $J$ in NRQCD since the hard contribution has been integrated out in the NRQCD effective theory~\cite{Marquard:2014pea,Bekavac:2009zc}.  
The lhs and rhs of the matching formula represent the renormalization of $\Gamma_{J}$ in QCD and $\widetilde{\Gamma}_{J}$ in NRQCD, respectively.

$Z_{2,b(c)}^\mathrm{OS}$ and $\widetilde{Z}_{2,b(c)}^\mathrm{OS}$  are the $b(c)$ quark field   renormalization constants in the on-shell ($\mathrm{OS}$) scheme in  QCD and NRQCD, respectively.
$Z_{2,b(c)}^\mathrm{OS}$ can be found in ref.~\cite{Fael:2020bgs} while 
$\widetilde{Z}_{2,b(c)}^\mathrm{OS}=1$ since heavy bottom and charm quarks are decoupled in the NRQCD effective theory~\cite{Grozin:2011nk,Bekavac:2009zc}.
${\widetilde Z}_{J}$ is the NRQCD  heavy flavor-changing  current  renormalization constant in the  modified-minimal-subtraction    (${\overline{\mathrm{MS}}}$) scheme.
${Z_{J}^\mathrm{OS}} (Z_{J}^{\overline{\mathrm{MS}}})$ is the QCD heavy flavor-changing current  renormalization constant in $\mathrm{OS}({\overline{\mathrm{MS}}}$) scheme.
If $Z_{J}^{\mathrm{OS}}$ is known, ${\widetilde Z}_{J}$ and $\mathcal{C}_{J}$  can be directly calculated by inputting $Z_{J}^{\mathrm{OS}}$  in eq.~\eqref{matchingformula}.
Similarly, if  $Z_{J}^{\overline{\mathrm{MS}}}$ is known, ${\widetilde Z}_{J}$ and  $\overline{\mathcal{C}}_{J}$ can be directly calculated by inputting  $Z_{J}^{\overline{\mathrm{MS}}}$ in eq.~\eqref{matchingformula}.
At the leading-order (LO) of $\alpha_s$ we set  $\mathcal{C}_{J}^{\text{LO}}=\overline{\mathcal{C}}_{J}^{\text{LO}}=1$, while at the higher-order of $\alpha_s$ both $\mathcal{C}_{J}$ and $\overline{\mathcal{C}}_{J}$ are  finite in $D=4$ and  depend on the NRQCD factorization scale $\mu_f$ and QCD renormalization scale $\mu$~\cite{Tao:2023vvf}.
Note that $Z_{J}^{\mathrm{OS}}$, $Z_{J}^{\overline{\mathrm{MS}}}$, ${\widetilde Z}_{J}$, $\mathcal{C}_{J}$ and $\overline{\mathcal{C}}_{J}$ are gauge invariant to all orders in $\alpha_s$~\cite{Tao:2023mtw,Tao:2023vvf}.

From the NRQCD currents in eq.~\eqref{NRQCDcurrents}, we can establish  the following relations   between  the NRQCD current renormalization constants, the matching coefficients and the decay constants associated with different heavy flavor-changing currents~\cite{Chetyrkin:1996ela,Lee:2018rgs,Fael:2022miw,Kniehl:2006qw,Bi:2017ybi,Blum:2001sr,Sturm:2009kb,Herdoiza:2006qv,McNeile:2006nv}:
\begin{align}\label{NRQCDrelations}
{\widetilde Z}_{s}={\widetilde Z}_{v,0},~~
&\frac{f_{B_{c0}^*}^s}{f_{B_{c0}^*}^{v,0}}=\frac{\mathcal{C}_s}{\mathcal{C}_{v,0}}=\frac{m_bZ_{m,b}^\mathrm{OS}+m_cZ_{m,c}^\mathrm{OS}}{m_bZ_{m,b}^\mathrm{OS}-m_cZ_{m,c}^\mathrm{OS}}\frac{m_b-m_c}{m_b+m_c};
\nonumber\\
&\frac{f_{B_c}^{p}}{f_{B_c}^{a,0}}=\frac{\mathcal{C}_{p}}{\mathcal{C}_{a,0}}=\frac{{\widetilde Z}_{p}}{{\widetilde Z}_{a,0}}=1;
\nonumber\\
&\frac{f_{B_c^*}^{v,i}}{f_{B_c^*}^{t,i0}}=\frac{\mathcal{C}_{v,i}}{\mathcal{C}_{t,i0}}=\frac{{\widetilde Z}_{v,i}}{{\widetilde Z}_{t,i0}}=1,
\end{align}
where $Z_{m,b(c)}^\mathrm{OS}$ is the $b(c)$ quark mass  $\mathrm{OS}$ renormalization constant in QCD,
which  can be obtained from ref.~\cite{Fael:2020bgs}.

Based on the matching formula  in eq.~\eqref{matchingformula}, we can also derive  the following relations among $Z_{J}^\mathrm{{OS}}$, $Z_{J}^\mathrm{\overline{MS}}$, $\mathcal{C}_{J}$ and $\overline{\mathcal{C}}_{J}$:
\begin{align}\label{QCDrelation0} 
&\frac{\mathcal{C}_{J}}{\overline{\mathcal{C}}_{J}}=\frac{Z_{J}^{\mathrm{OS}}}{Z_{J}^{\mathrm{\overline{MS}}}}=z_J^g z_J^\mu+{\mathcal O}(\epsilon),
\end{align}
which, combined with eq.~\eqref{NRQCDrelations}, yield~\cite{Bauer:2000yr}:
\begin{align}
J\in&\{s,p\}:
\nonumber\\&
Z_{J}^\mathrm{\overline{MS}}=Z_m^\mathrm{\overline{MS}},
\\&
Z_{J}^\mathrm{OS}=\frac{m_b Z_{m,b}^\mathrm{OS}+m_c Z_{m,c}^\mathrm{OS}}{m_b+m_c},
\\&
z_J^g z_J^\mu=z_m^g z_m^\mu=\frac{\mathcal{C}_{a,0}}{\overline{\mathcal{C}}_{p}},
\\&
{\mathcal{C}}_{J}=z_m^g z_m^\mu \overline{\mathcal{C}}_{J};
\\
J\in&\{(v,0),(v,i),(a,0),(a,i)\}:
\nonumber\\&
Z_{J}^{\mathrm{\overline{MS}}}=Z_{J}^{\mathrm{OS}}=z_J^g z_J^\mu=1,
\\&
{\mathcal{C}}_{J}= \overline{\mathcal{C}}_{J};
\\
J\in&\{(t,i0),(t,ij),(t5,i0),(t5,ij)\}:
\nonumber\\&
Z_{J}^\mathrm{\overline{MS}}=Z_{t}^\mathrm{\overline{MS}},
\\&
Z_{J}^\mathrm{OS}=Z_{t}^\mathrm{OS},
\\&
z_J^g z_J^\mu=z_t^g z_t^\mu=\frac{\mathcal{C}_{v,i}}{\overline{\mathcal{C}}_{t,i0}},\label{ztgztmu}
\\&
{\mathcal{C}}_{J}=z_t^g z_t^\mu \overline{\mathcal{C}}_{J}.\label{Cbar2C}
\end{align}
$Z_m^\mathrm{\overline{MS}}$ is the quark mass $\mathrm{\overline{MS}}$ renormalization constant in QCD, which can be found in refs.~\cite{Marquard:2016dcn,Gracey:2000am,Broadhurst:1994se,Bell:2010mg,Blumlein:2018tmz}.
$Z_t^{\mathrm{\overline{MS}}}$ is the QCD tensor current $\mathrm{\overline{MS}}$ renormalization constant, which %can be obtained from refs.
is available in various literature~\cite{Gracey:2000am,Broadhurst:1994se,Bell:2010mg,Baikov:2006ai,Gracey:2022vqr,Tao:2023vvf}.
$Z_t^{\mathrm{OS}}$ is the QCD heavy flavor-changing tensor current ${\mathrm{OS}}$ renormalization constant. 
Although the complete three-loop expression of $Z_t^{\mathrm{OS}}$ is   
%not available in the literature,
unknown at present,
the finite ($\epsilon^0$) term $z_t^gz_t^\mu$ of the ratio $Z_t^{\mathrm{OS}}/Z_t^{\mathrm{\overline{MS}}}$ can be calculated by eq.~\eqref{ztgztmu} and the three-loop result of $z_t^gz_t^\mu$  has been 
%known    in our earlier paper~\cite{Tao:2023vvf}. 
obtained by our earlier work~\cite{Tao:2023vvf}.
For  $J\in\{(t,i0),(t,ij),(t5,i0),(t5,ij)\}$, in order to calculate the matching coefficient ${\mathcal{C}}_{J}$, 
we firstly calculate ${\widetilde Z}_{J}$ and $\overline{\mathcal{C}}_{J}$ by inputting $Z_{J}^{\overline{\mathrm{MS}}}$ in eq.~\eqref{matchingformula}, then ${\mathcal{C}}_{J}$
can be obtained by eq.~\eqref{Cbar2C}.

\section{QCD vertex function~\label{QCDvertexfunction}}

In this section, we will present the details of our calculations  for the QCD vertex functions with the heavy flavor-changing temporal-vector $(v,0)$, spatial-spatial tensor $(t,ij)$ and spatial-temporal axial-tensor $(t5,i0)$ currents, which are coupled to the $P$-wave $c\bar b$ mesons $B_{c0}^*$ and $B_{c1}$. 
And the calculation details for the QCD vertex functions with other seven currents can be found in our previous  research~\cite{Tao:2022hos,Tao:2023mtw,Tao:2023vvf}.

\begin{figure}[htbp]
	\center{
		\includegraphics*[scale=0.7]{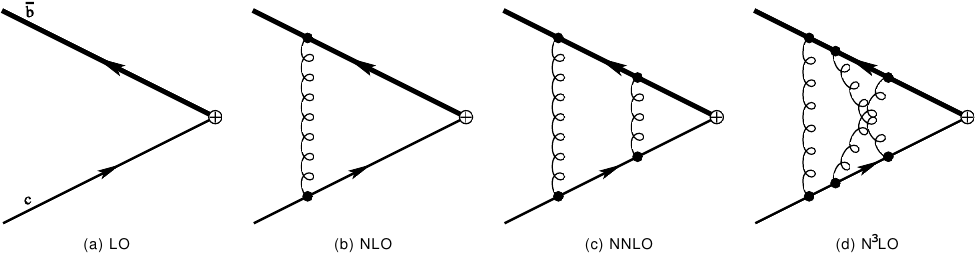}%\qquad
		\caption {\label{fig:picupto3loop} Representative Feynman diagrams up to three loops	for the QCD vertex function with the heavy flavor-changing   current involving bottom and charm quarks.	 The cross ``$\bigoplus$'' implies the insertion of a certain heavy flavor-changing   current.	}}
\end{figure}
The sample Feynman diagrams  contributing to the QCD vertex function are displayed in figure~\ref{fig:picupto3loop}.
The on-shell amputated  QCD amplitudes with tensor structures for the vector, tensor and  axial-tensor currents
can be  denoted by  $\Gamma_{(v)}^{\mu}=\cdots\gamma^{\mu}\cdots$,  $\Gamma_{(t)}^{\mu\nu}=\cdots\sigma^{\mu\nu}\cdots$, $\Gamma_{(t5)}^{\mu\nu}=\cdots\sigma^{\mu\nu}\gamma_5\cdots$, respectively. 
We denote the  external momenta of charm and bottom quarks by  $q_c$ and $q_b$, respectively,
which also constitute the complete external momenta of the   QCD amplitudes.
Let $q=q_c+q_b$ represent the total external momentum and the small momentum $k$ refer to the relative movement  between the bottom and charm quarks. 
In order to obtain the hard QCD vertex function $\Gamma_{J}$ with $J\in\{(v,0),(t,ij),(t5,i0)\}$, we need to first replace $q_c$ and $q_b$ with $q$ and $k$, and then  calculate the contributions from terms up to ${\mathcal{O}(k)}$ (see eq.~\eqref{NRQCDcurrents}) in the   QCD amplitude. 
Choosing a reference frame where   $q\cdot k=0$, from the on-shell condition $q_c^2=m_c^2,q_b^2=m_b^2$   we can  derive the  following external momentum configuration~\cite{Zhu:2017lqu}:
\begin{align}
&q_c=\frac{\sqrt{m_c^2-k^2}}{\sqrt{m_b^2-k^2}+\sqrt{m_c^2-k^2}}\,q+k,
\nonumber\\&
q_b=\frac{\sqrt{m_b^2-k^2}}{\sqrt{m_b^2-k^2}+\sqrt{m_c^2-k^2}}\,q-k;
\nonumber\\
&q^2=\left(\sqrt{m_b^2-\delta}+\sqrt{m_c^2-\delta}\right)^2,
\nonumber\\&
q\cdot k=0,
~~~
k^2=\delta,
\end{align}
where  $\delta$ is a small quantity.
After  removing the non-contributing terms of $\mathcal{O}{(k^n)}(n\geq 2)$ from the rhs of the first three equations above,  the external momentum configuration can be simplified as
\begin{align}\label{mmconfiguration}
&q_c=\frac{m_c}{m_b+m_c}q+k,
\nonumber\\
&q_b=\frac{m_b}{m_b+m_c}q-k;
\nonumber\\
&q^2=(m_b+m_c)^2,
\nonumber\\&
q\cdot k=0,
~~~
k^2=\delta.
\end{align}

%As previously mentioned, the $\mathcal{O}{(k)}$ terms in the QCD amplitude are needed to determine the hard QCD vertex function, which requires an expansion of the QCD amplitude with respect to  $k$ up to $\mathcal{O}{(k)}$ in the hard integration region. With the external momentum configuration in eq.~\eqref{mmconfiguration}, the QCD amplitude contains propagators involving the small momentum $k$. Consequently,  each propagator  needs to be expanded in a power series of $k$ up to $\mathcal{O}{(k)}$ in the hard region of loop momenta.

Within the external momentum configuration in eq.~\eqref{mmconfiguration}, the QCD amplitude contains propagators involving the small momentum $k$. Since the $\mathcal{O}{(k)}$ terms in the QCD amplitude are needed to determine the hard QCD vertex function, it is required to expand the QCD amplitude with respect to $k$ up to $\mathcal{O}{(k)}$ in the hard integration region, which means each propagator needs to be expanded in a power series of $k$ up to $\mathcal{O}{(k)}$ in the hard region of loop momenta.

Following the literature~\cite{Onishchenko:2003ui,Kniehl:2006qw}, we employ the  projector on the QCD amplitude to extract the hard QCD vertex function. The projectors for the currents $(v,0),(t,ij),(t5,i0)$ can be constructed as
\begin{small}
\begin{align}
P_{(v,0),\mu} =\,& \frac{1}{2(m_b+m_c)^2} \bigg\{
\frac{m_c}{m_c-m_b}\left(\frac{m_c}{m_b+m_c}\slashed{q} + m_c
\right) \gamma_{\mu}\left(\frac{m_b}{m_b+m_c}\slashed{q} + m_b
\right)
\nonumber\\&
-\frac{m_b}{m_c-m_b}\left(-\frac{m_c}{m_b+m_c}\slashed{q} + m_c
\right) \gamma_{\mu}\left(-\frac{m_b}{m_b+m_c}\slashed{q} + m_b
\right)
\nonumber\\&
+ \frac{2 m_b m_c}{m_c-m_b} \left(\frac{m_c}{m_b+m_c}\slashed{q} + m_c
\right) \frac{\gamma_{\mu} \slashed{k} }{k^2}
\left(-\frac{m_b}{m_b+m_c}\slashed{q} + m_b\right) \bigg\},
\nonumber\\
P_{(t,ij),\mu\nu} =\,& \frac{1}{2(D-2)(m_b+m_c)^2} \bigg\{
\frac{1}{D-1}\frac{m_c}{m_b+m_c}\left(\frac{m_c}{m_b+m_c}\slashed{q} + m_c
\right) \sigma_{\mu\nu}\left(\frac{m_b}{m_b+m_c}\slashed{q} + m_b
\right)
\nonumber\\&
+\frac{1}{D-1}\frac{m_b}{m_b+m_c}\left(-\frac{m_c}{m_b+m_c}\slashed{q} + m_c
\right) \sigma_{\mu\nu}\left(-\frac{m_b}{m_b+m_c}\slashed{q} + m_b
\right)
\nonumber\\&
+ \frac{1}{D-3}\frac{2 m_b m_c}{m_b+m_c} \left(\frac{m_c}{m_b+m_c}\slashed{q} + m_c
\right) \frac{2 {\rm i}k_\mu \gamma_\nu + \sigma_{\mu\nu} \slashed{k} }{k^2}
\left(-\frac{m_b}{m_b+m_c}\slashed{q} + m_b\right) \bigg\},
\nonumber\\
P_{(t5,i0),\mu\nu} =\, & \frac{1}{4(m_b+m_c)^2} \bigg\{
\frac{1}{D-1}\frac{m_c}{m_b+m_c}\left(\frac{m_c}{m_b+m_c}\slashed{q} + m_c
\right) \sigma_{\mu\nu}\gamma_5\left(\frac{m_b}{m_b+m_c}\slashed{q} + m_b
\right)
\nonumber\\&
+\frac{1}{D-1}\frac{m_b}{m_b+m_c}\left(-\frac{m_c}{m_b+m_c}\slashed{q} + m_c
\right) \sigma_{\mu\nu}\gamma_5\left(-\frac{m_b}{m_b+m_c}\slashed{q} + m_b
\right)
\nonumber\\&
+ \frac{4 m_b m_c}{m_b+m_c} \left(\frac{m_c}{m_b+m_c}\slashed{q} + m_c
\right) \frac{ {\rm i}k_\mu \gamma_\nu \gamma_5 }{k^2}
\left(-\frac{m_b}{m_b+m_c}\slashed{q} + m_b\right) \bigg\}.
\end{align}
\end{small}

By applying the projectors to the expanded QCD amplitudes, the hard QCD vertex functions  can be obtained as following
\begin{align}
%\Gamma_{t,i0} = &\mbox{Tr}\left[ {P_{(t,i0),\mu\nu} \Gamma_{(t)}^{\mu\nu}}_{|k=0} \right]\,,
%\nonumber\\
%\Gamma_{t5,ij} = &\mbox{Tr}\left[ {P_{(t5,ij),\mu\nu} \Gamma_{(t5)}^{\mu\nu}}_{|k=0} \right]\,,
%\nonumber\\
\Gamma_{v,0} &= \lim_{\delta\to 0}\mbox{Tr}\left[ P_{(v,0),\mu} \Gamma_{(v)}^{\mu} \right]\,,
\nonumber\\
\Gamma_{t,ij} &=\lim_{\delta\to 0} \mbox{Tr}\left[ P_{(t,ij),\mu\nu} \Gamma_{(t)}^{\mu\nu} \right]\,,
\nonumber\\
\Gamma_{t5,i0} &=\lim_{\delta\to 0} \mbox{Tr}\left[ P_{(t5,i0),\mu\nu} \Gamma_{(t5)}^{\mu\nu} \right]\,.
\end{align}
We want to mention that we employ the  naively anticommuting $\gamma_5$ scheme throughout the calculations for matching coefficients involving heavy flavor-changing currents due to the absence of traces with an odd number of $\gamma_5$~\cite{Piclum:2007an,Tao:2023mtw}.

For the multi-loop calculation of matching coefficients of currents with $P$-wave $c\bar b$ mesons, we utilize the same automated setup as that with $S$-wave   $c\bar b$ mesons described in ref.~\cite{Tao:2023vvf}, while also  introducing two new packages {\texttt{FormLink}~\cite{Feng:2012tk}} and {\texttt{Fermatica}~\cite{Lee:2020zfb}}
to deal with the large expressions in the Feynman amplitudes. 
Firstly, we  use {\texttt{FeynCalc}~\cite{Shtabovenko:2020gxv}} + {\texttt{FormLink}}  to obtain Feynman diagrams and corresponding Feynman amplitudes. 
There are 1, 1, 13, 268 bare  Feynman diagrams at  tree, one-loop, two-loop, three-loop orders, respectively,   contributing to the QCD vertex function with each heavy flavor-changing   current.
By {\texttt{\$Apart}~\cite{Feng:2012iq}} + {\texttt{Fermatica}}, each Feyman amplitude is decomposed into several families of Feynman integrals. 
In each family of three-loop Feynman integrals, there are twelve inverse propagators where nine are free from the small momentum $k$ and the remaining three are irreducible numerators involving $k$.
Based on the symmetry among different families, we use   {our \texttt{Mathematica} code} + {\texttt{LiteRed}~\cite{Lee:2013mka}} + {\texttt{FIRE6}~\cite{Smirnov:2019qkx}}  to  minimize~\cite{Fael:2020njb,Shtabovenko:2021hjx,Gerlach:2022qnc}  the total number of families for all Feynman integrals. 
For each current involving the $P$-wave $c\bar b$ meson, the total number of families for  three-loop Feynman integrals is minimized from 841 to 102.
Then, we use {\texttt{FIRE6}}/{\texttt{Kira}~\cite{Klappert:2020nbg}}/{\texttt{FiniteFlow}~\cite{Peraro:2019svx}} to reduce each family of Feynman integrals to the family of master integrals  based on Integration by Parts (IBP)~\cite{Chetyrkin:1981qh}.
Next, we use {our \texttt{Mathematica} code} + {\texttt{Kira}} + {\texttt{FIRE6}}  to  minimize the total  number of families for all   master integrals.
For each current involving the $P$-wave $c\bar b$ meson, the total number of families for  three-loop master integrals  is minimized from 102 to 26, with 
these 26 families collectively containing a total of 300 three-loop master integrals.
%with the latter composed of 300 three-loop matser integrals.
%After minimizing 102 three-loop families for each current involving the $P$-wave $c\bar b$ meson, we end with 26 families, totaling  300 three-loop master integrals.
Last, we use {\texttt{AMFlow}~\cite{Liu:2022chg}}, which is a proof-of-concept implementation of the auxiliary mass flow method~\cite{Liu:2017jxz,Liu:2021wks,Liu:2022mfb}, equipped with {\texttt{FiniteFlow}/\texttt{Kira}} to calculate each family of master integrals.

\section{NRQCD current renormalization constant~\label{ZjNRQCD}}

The bare hard QCD vertex function $\Gamma_{J}$ can be renormalized~\cite{Davydychev:1997vh,Bekavac:2009zc} by the lhs of eq.~\eqref{matchingformula}, which involves the QCD heavy quark field and  mass ${\rm OS}$ renormalization,  the QCD coupling constant $\overline{\rm MS}$ renormalization~\cite{Mitov:2006xs,Chetyrkin:1997un,vanRitbergen:1997va}, 
as well as the QCD heavy flavor-changing current ${\rm OS}(\overline{\rm MS})$ renormalization.
After implementing %the aforementioned
these renormalization procedures, the lhs of eq.\eqref{matchingformula} still contains uncancelled infra-red poles starting from order $\alpha_s^2$, which completely determine the poles of the NRQCD current $\overline{\rm MS}$ renormalization constant ${\widetilde Z}_{J}$ on the rhs of eq.\eqref{matchingformula} by rendering ${\mathcal{C}_J(\overline{\mathcal{C}_J})}$ finite in $D=4$~\cite{Bekavac:2009zc}.

With the aid of numerical fitting techniques such as the PSLQ algorithm~\cite{Duhr:2019tlz}, we have individually reconstructed the analytic expressions of  ${\widetilde Z}_{J}$ for all ten heavy flavor-changing currents, where ${\widetilde Z}_{J}$ for  $J\in\{(v,0),(t,ij),(t5,i0)\}$ are new while ${\widetilde Z}_{J}$ for  the remaining seven currents are already known  
%well-documented
in previous literature~\cite{Feng:2022ruy,Sang:2022tnh,Tao:2022hos,Tao:2023mtw,Tao:2023vvf}.
Our results verify ${\widetilde Z}_{s}\equiv{\widetilde Z}_{v,0}$ and the other two relations for ${\widetilde Z}_{J}$  as described in eq.~\eqref{NRQCDrelations}, thus the result of ${\widetilde Z}_{v,0}$ can be obtained directly from the result of ${\widetilde Z}_{s}$   and will not be repeated here.
Since all of the  currents $(a,i)$, $(t,ij)$, and $(t5,i0)$ can couple to (can produce/annihilate) the $P$-wave axial-vector  $B_{c1}$ meson, we will present the results of ${\widetilde Z}_{t,ij}$ and ${\widetilde Z}_{t5,i0}$, along with ${\widetilde Z}_{a,i}$,  for the purpose of convenient comparison.
Throughout the paper, we define the following dimensionless variables:
\begin{align}
x&\equiv {m_c\over m_b},
\nonumber\\
L_{\mu}&\equiv \ln \frac{\mu^2}{m_b m_c},
\nonumber\\
L_{\mu_f}&\equiv \ln \frac{\mu_f^2}{m_b m_c}.
\end{align}
Then ${\widetilde Z}_{J}$ for  $J\in\{(a,i),(t,ij),(t5,i0)\}$ can be expressed as
\begin{align}
\widetilde{Z}_{J}
=\,&1+\sum_{i=1}^{\infty}\left(\frac{\alpha_{s}^{(n_{l})}(\mu_f)}{\pi}\right)^{i}\widetilde{Z}_{J}^{(i)},
\nonumber\\
\widetilde{Z}_{J}^{(1)}=\,&0;
\nonumber \\
\widetilde{Z}_{J}^{(2)}=\,&\pi^{2}C_{F}\frac{1}{\epsilon}\left(\frac{3x^2+c_1^J\,x+3}{24\left(1+x\right)^2}C_{F}+\frac{1}{24}C_{A}\right),
\nonumber\\
\widetilde{Z}_{J}^{(3)}=\,&
\pi^{2}C_{F}\bigg\{
C_F^2 \frac{1}{\epsilon}\left(
\frac{57 x^2+c_2^J\, x+57}{216 (x+1)^2}-\frac{1 }{3}\ln 2
\right.
%\nonumber\\&
\left.
+\frac{c_3^J\,x(x-1)}{48 (x+1)^3}\ln x
\right)
\nonumber\\ &
+
C_F C_A \bigg[
-\frac{22x^2+c_4^J\,x+22}{432\epsilon ^2 (x+1)^2 }
+
\frac{1}{ \epsilon }\left(
\frac{379 x^2+c_5^J\,	x+379}{1296 (x+1)^2}-\frac{1}{18} \ln 2
\right.\nonumber\\&\left.
-\frac{5x+11 }{144 (x+1)}\ln x
+\frac{1}{9} \ln (x+1)
+\frac{11 x^2+c_6^J\, x+11}{144 (x+1)^2} L_{\mu _f}
\right)\bigg]
\nonumber\\ &
+C_A^2  \bigg[
\frac{-1}{48 \epsilon ^2}+\frac{1}{ \epsilon
}\left(
\frac{17}{324}+\frac{1}{9}\ln2-\frac{1}{72}\ln x+\frac{1}{36}\ln (x+1) +
\frac{1}{72}L_{\mu _f}
\right)\bigg]
\nonumber\\ &
+ C_F T_F n_l\bigg[
\frac{3 x^2+c_7^J\,
	x+3}{108\epsilon ^2 (x+1)^2 }-\frac{21 x^2+c_8^J\, x+21}{324 \epsilon (x+1)^2 }
\bigg]
\nonumber\\&
+C_A  T_F n_l \bigg[
\frac{1}{108 \epsilon ^2}-\frac{53}{1296 \epsilon }
\bigg]
\bigg\},
\end{align}
with
\begin{align}
&c_1^{a,i}=4,\,c_1^{t,ij}=c_1^{t5,i0}=2;
\nonumber\\&
c_2^{a,i}=101,\,c_2^{t,ij}=c_2^{t5,i0}=92;
\nonumber\\&
c_3^{a,i}=1,\,c_3^{t,ij}=c_3^{t5,i0}=4;
\nonumber\\&
c_4^{a,i}=31,\,c_4^{t,ij}=c_4^{t5,i0}=12;
\nonumber\\&
c_5^{a,i}=675,\,c_5^{t,ij}=502,\,c_5^{t5,i0}=484;
\nonumber\\&
c_6^{a,i}=13,\,c_6^{t,ij}=c_6^{t5,i0}=10;
\nonumber\\&
c_7^{a,i}=4,\,c_7^{t,ij}=c_7^{t5,i0}=2;
\nonumber\\&
c_8^{a,i}=41,\,c_8^{t,ij}=c_8^{t5,i0}=22,
\end{align}
where $n_l$ is the number of massless flavors.
The corresponding NRQCD heavy flavor-changing   current anomalous dimension $\tilde{\gamma}_{J}$   is related to  $\widetilde{Z}_{J}$ by~\cite{Groote:1996xb,Kiselev:1998wb,Henn:2016tyf,Fael:2022miw,Grozin:2015kna,Ozcelik:2021zqt}
\begin{align}
&\tilde{\gamma}_{J} =\sum_{i=1}^{\infty}\left(\frac{\alpha_s^{(n_l)}
	\left(\mu_f\right)}{\pi}\right)^i \tilde{\gamma}_{J}^{(i)},
\nonumber\\
&\tilde{\gamma}_{J}^{(i)}=-2\,i\,\widetilde{Z}_{J}^{(i)[1]},
\end{align}
where  $\widetilde{Z}_{J}^{(i)[1]}$ denotes the coefficient of  $\frac{1}{\epsilon}$  in $\widetilde{Z}_{J}^{(i)}$. 
Note that both $\widetilde{Z}_{J}$ and $\tilde{\gamma}_{J}$  are invariant   under the exchange $m_b\leftrightarrow m_c$ meanwhile $n_b\leftrightarrow n_c$.

In QCD we consider $n_l$ massless quarks,  $n_b$ bottom quarks with mass $m_b$ and $n_c$ charm quarks  with mass $m_c$ possibly appearing in the quark loop, 
while the heavy charm  and bottom quarks  are decoupled in the NRQCD.  
For matching between QCD and NRQCD with the same perturbative expansion parameter $\alpha_s$, 
we employ both the coupling running~\cite{Abreu:2022cco,Tao:2022hos,Tao:2023mtw} and the decoupling relation~\cite{Chetyrkin:2005ia,Kniehl:2006bg,Bernreuther:1981sg,Grozin:2011nk,Grozin:2012ic,Barnreuther:2013qvf,Grozin:2007fh,Gerlach:2019kfo,Ozcelik:2021zqt,Tao:2022hos,Tao:2023mtw} in $D=4-2\epsilon$ 
for the  interconversion among 
$\alpha_s^{(n_f)}(\mu)$, $\alpha_s^{(n_l)}(\mu_f)$ and $\alpha_s^{(n_l)}(\mu)$, where
$n_f=n_l+n_b+n_c$ is the total number of flavors, 
$n_l$ is the number of massless flavors,  $n_{b(c)}$ is the number of bottom (charm) flavors with mass $m_{b(c)}$, $\mu$ is the QCD renormalization scale and $\mu_f$ is the NRQCD factorization scale.
To compute the numerical value of $\alpha_s^{(n_l)}(\mu)$ in $D=4$ with $n_l=3,\,n_b=n_c=1$ and $\mu\in[0.4,7]\,\mathrm{GeV}$, 
we can directly  utilize the function {\texttt{AlphasLam}} in   the package  {\texttt{RunDec}~\cite{Chetyrkin:2000yt,Schmidt:2012az,Deur:2016tte,Herren:2017osy}}     with  $\Lambda_{QCD}^{(n_l=3)}=0.3344\mathrm{GeV}$ 
determined  at  three-loop accuracy  by inputting the initial value  $\alpha_s^{(n_f=5)}\left(m_Z=91.1876\mathrm{GeV}\right)=0.1179$.

\section{Matching coefficient~\label{Matchingcoefficient}}

The matching coefficient $\mathcal{C}_J$ for the heavy flavor-changing current 
$J$ in eq.~\eqref{j10}
can be expressed in the following form~\cite{Feng:2022ruy,Sang:2022tnh,Tao:2022hos,Tao:2023mtw,Tao:2023vvf}:
\begin{align}
\label{Cjformula}
&\mathcal{C}_J(\mu_f,\mu,m_b,m_c) = 1+\frac{\alpha_s^{(n_l)}(\mu)}{\pi} \mathcal{C}_J^{(1)}(x)
\nonumber \\&
+\left(\frac{\alpha_s^{(n_l)}(\mu)}{\pi}\right)^2
\bigg[\frac{\mathcal{C}_J^{(1)}(x)}{4}\beta_0^{(n_l)}L_{\mu }
+\frac{\tilde{\gamma}_J^{(2)}(x)}{2}L_{\mu _f}+\mathcal{C}_J^{(2)}(x)\bigg]
\nonumber \\&
 +\left(\frac{\alpha_s^{(n_l)}(\mu)}{\pi}\right)^3\Bigg\lbrace
\frac{\mathcal{C}_J^{(1)}(x)}{16}{\beta_0^{(n_l)}}^2L^2_{\mu}
+\bigg[\frac{\mathcal{C}_J^{(1)}(x)}{16}\beta_1^{(n_l)}+\frac{\mathcal{C}_J^{(2)}(x)}{2}\beta_0^{(n_l)}\bigg]L_{\mu }
\nonumber \\&
+\frac{\tilde{\gamma}_J^{(2)}(x)}{4}\beta_{0}^{(n_l)}L_{\mu }L_{\mu _f}
+\bigg[\frac{\partial\tilde{\gamma}_J^{(3)}\left(L_{\mu _f};x\right)}{4\partial L_{\mu _f}}-\frac{\tilde{\gamma}_J^{(2)}(x)}{8}\beta_0^{(n_l)}\bigg]L^2_{\mu _f} 
\nonumber \\&
 +\frac{1}{2}\bigg[\mathcal{C}_J^{(1)}(x) \tilde{\gamma}_J^{(2)}(x)+\tilde{\gamma}_J^{(3)}\left(L_{\mu _f}=0;x\right)\bigg]L_{\mu _f}
+ \mathcal{C}_J^{(3)}(x) \Bigg\rbrace+\mathcal{O}\left(\alpha_s^4\right),
\end{align}
where  $\beta_0^{(n_l)}=\frac{11}{3}C_A-\frac{4}{3} T_F n_l$ and  $\beta_1^{(n_l)}=\frac{34}{3}C_A^2-4 C_F T_F n_l-\frac{20}{3} C_A T_F n_l$ correspond to the one-loop and two-loop coefficients of the QCD $\beta$ function~\cite{vanRitbergen:1997va}, respectively, and the undetermined functions $\mathcal{C}_J^{(n)}(x)\,(n=1,2,3)$ for the current $J$  solely depend on $x=m_c/m_b$.

The analytic results of $\mathcal{C}_J^{(1)}(x)$ for various currents read:
\begin{align}
&\mathcal{C}_{s}^{(1)}\left({x}\right)=\frac{3}{4}C_F\left(\frac{x-1}{x+1}\,\ln x-\frac{2}{3}\right),
\nonumber\\&
\mathcal{C}_{v,0}^{(1)}(x)=\frac{3}{4}C_F\left(\frac{x+1}{x-1}\,\ln x-\frac{2}{3}\right),
\nonumber\\&
\mathcal{C}_p^{(1)}(x)=\mathcal{C}_{a,0}^{(1)}(x)=\frac{3}{4}C_F\left(\frac{x-1}{x+1}\,\ln x-2\right),
\nonumber\\&
\mathcal{C}_{v,i}^{(1)}(x)=\mathcal{C}_{t,i0}^{(1)}(x)=\mathcal{C}_{t5,ij}^{(1)}(x)=\frac{3}{4}C_F\left(\frac{x-1}{x+1}\,\ln x-\frac{8}{3}\right),
\nonumber\\&
\mathcal{C}_{a,i}^{(1)}(x)=\mathcal{C}_{t,ij}^{(1)}(x)=\mathcal{C}_{t5,i0}^{(1)}(x)=\frac{3}{4} C_F \left(\frac{x-1}{x+1}\,\ln x-\frac{4}{3}\right).
\end{align}

$\mathcal{C}_J^{(2)}(x)$ and $\mathcal{C}_J^{(3)}(x)$ can be decomposed according to the different color factors~\cite{Fael:2020bgs,Marquard:2014pea,Beneke:2014qea,Egner:2022jot,Feng:2022vvk,Feng:2022ruy,Sang:2022tnh,Tao:2022hos,Tao:2023mtw,Tao:2023vvf}:
\begin{align}
\mathcal{C}_J^{(2)}(x) =\,& C_F \,\Big[ C_F \,\mathcal{C}_J^{FF}(x)+ C_A\, \mathcal{C}_J^{FA}(x) 
% \nonumber\\&
+ T_F\, n_l\, \mathcal{C}_J^{FL}(x)+ T_F\, n_b \,\mathcal{C}_J^{FB}(x) + T_F\, n_c\, \mathcal{C}_J^{FC}(x)\Big];
\nonumber\\
\mathcal{C}_J^{(3)}(x) =\,& C_F\,\Big[ C^2_F \, \mathcal{C}_J^{FFF}(x)+C_F \,C_A \, \mathcal{C}_J^{FFA}(x)
+ C_A^2 \, \mathcal{C}_J^{FAA}(x)
\nonumber\\&
+C_F\, T_F\, n_l\, \mathcal{C}_J^{FFL}(x)+ C_F\,  T_F\, n_b\, \mathcal{C}_J^{FFB}(x)+
C_F\,T_F \, n_c\,  \mathcal{C}_J^{FFC}(x) 
\nonumber\\&
+C_A\,T_F\, n_l\,\mathcal{C}_J^{FAL}(x) + C_A\,T_F\, n_b\, \mathcal{C}_J^{FAB}(x)  + C_A\,T_F \, n_c\, \mathcal{C}_J^{FAC}(x)
\nonumber\\&
+T_F^2\, n_l^2\, \mathcal{C}_J^{FLL}(x)+T_F^2 \, n_l \, n_b\,  \mathcal{C}_J^{FLB}(x)+T_F^2 \, n_l \, n_c\,  \mathcal{C}_J^{FLC}(x)  
\nonumber\\&
+T_F^2\, n_b^2\,
\mathcal{C}_J^{FBB}(x)+ T_F^2 \, n_b \, n_c \, \mathcal{C}_J^{FBC}(x) +
T_F^2 \, n_c^2\,  \mathcal{C}_J^{FCC}(x)   
\Big]. 
\end{align}

In the forthcoming, we will present the numerical results with about 30-digit precision for the color-factor components of $\mathcal{C}_J^{(2)}(x)$ and  $\mathcal{C}_J^{(3)}(x)$
with $J\in\{(v,0),(t,ij),(t5,i0)\}$ at the physical heavy quark mass ratio  $x=x_0=\frac{150}{475}$, which are calculated for the first time, while the results of the matching coefficients for the other seven currents can be found in various literature~\cite{Feng:2022ruy,Sang:2022tnh,Tao:2022hos,Tao:2023mtw,Tao:2023vvf}.
The  color-factor components of ${\cal C}_{v,0}^{(2)}(x_0)$ and  ${\cal C}_{v,0}^{(3)}(x_0)$  read:
	\begin{align}
	\mathcal{C}_{v,0}^{FF}(x_0) &= -7.415072441701102157778077404885,
\nonumber	\\
	\mathcal{C}_{v,0}^{FA}(x_0) &=  -1.013040144406342708770557198762,
\nonumber	\\
	\mathcal{C}_{v,0}^{FL}(x_0) &= -1.239697820880047002421594209088,
\nonumber	\\
	\mathcal{C}_{v,0}^{FB}(x_0) &= 0.13597105048632699149910964134530,
\nonumber	\\
	\mathcal{C}_{v,0}^{FC}(x_0) &= 0.59363036585194212056527385664185;
\nonumber\\	
 \mathcal{C}_{v,0}^{FFF}(x_0) &=  -19.637407853517453632750614757579,
\nonumber	\\
	\mathcal{C}_{v,0}^{FFA}(x_0) &= -97.09492210039645832286078838246 ,
\nonumber	\\
	\mathcal{C}_{v,0}^{FAA}(x_0) &=  -53.06158531076542402693424442672,
\nonumber	\\
	\mathcal{C}_{v,0}^{FFL}(x_0) &= 30.31567533403874005372202227666,
\nonumber	\\
	\mathcal{C}_{v,0}^{FFB}(x_0) &= 1.274051979360092691782418978830,
\nonumber	\\
	\mathcal{C}_{v,0}^{FFC}(x_0) &= 4.991962080674365020786531737796 ,
\nonumber	\\
	\mathcal{C}_{v,0}^{FAL}(x_0) &= 8.297022736977196135197531673957,
\nonumber	\\
	\mathcal{C}_{v,0}^{FAB}(x_0) &= -0.575041406899846351438794824352,
\nonumber	\\
	\mathcal{C}_{v,0}^{FAC}(x_0) &= -0.678379635547194791872964103008 ,
\nonumber		\\
	\mathcal{C}_{v,0}^{FLL}(x_0) &= 1.952161757703498381986383836879 ,
\nonumber	\\
	\mathcal{C}_{v,0}^{FLB}(x_0) &= 0.07774011416568337646677489899643 ,
\nonumber	\\
	\mathcal{C}_{v,0}^{FLC}(x_0) &= -0.1377250733104515782404626852836 ,
\nonumber		\\
	\mathcal{C}_{v,0}^{FBB}(x_0) &= 0.04624649197909238074861442249224 ,
\nonumber	\\
	\mathcal{C}_{v,0}^{FBC}(x_0) &= 0.22322012566372742011700221329966 ,
\nonumber	\\
	\mathcal{C}_{v,0}^{FCC}(x_0) &= 0.3241345651384630564351480826585 .
	\end{align}
One can check that the  results of ${\mathcal C}_{v,0}$ above and those of ${\mathcal C}_{s}$ in refs.~\cite{Tao:2022hos,Tao:2023mtw} verify the relation for them in eq.~\eqref{NRQCDrelations}.

For the convenience of comparison, we present  the  color-factor components of ${\cal C}_{J}^{(2)}(x_0)$ and  ${\cal C}_{J}^{(3)}(x_0)$ for $J=(t,ij),(t5,i0)$,  along with $(a,i)$:
	\begin{align}
	\mathcal{C}_{a,i}^{FF}(x_0) &=-7.5581010098551732852894452744496 ,
\nonumber	\\
	\mathcal{C}_{t,ij}^{FF}(x_0) =\mathcal{C}_{t5,i0}^{FF}(x_0) &= -7.8095449630120296481644060237786,
\nonumber	\\
	\mathcal{C}_{a,i}^{FA}(x_0) &= -4.00029433499832597188660997836677,
\nonumber		\\
	\mathcal{C}_{t,ij}^{FA}(x_0) =\mathcal{C}_{t5,i0}^{FA}(x_0) &= -3.85213632317250612379027205712612,
\nonumber	\\
	\mathcal{C}_{a,i}^{FL}(x_0) =\mathcal{C}_{t,ij}^{FL}(x_0) =\mathcal{C}_{t5,i0}^{FL}(x_0) &= 0.064217493589576984548527284851041,
\nonumber	\\
	\mathcal{C}_{a,i}^{FB}(x_0) =\mathcal{C}_{t,ij}^{FB}(x_0) =\mathcal{C}_{t5,i0}^{FB}(x_0) &= 0.0140201311654815666291896168690850,
\nonumber	\\
	\mathcal{C}_{a,i}^{FC}(x_0) =\mathcal{C}_{t,ij}^{FC}(x_0) =\mathcal{C}_{t5,i0}^{FC}(x_0) &= 0.1451892135460458985527809307761497;
\nonumber	\\	
	\mathcal{C}_{a,i}^{FFF}(x_0) &= -8.42643917300377651280522219395,
	\nonumber	\\	
	\mathcal{C}_{t,ij}^{FFF}(x_0) &=-0.928357583746980594010510379717 ,
\nonumber		\\	
	\mathcal{C}_{t5,i0}^{FFF}(x_0) &= -0.598500030847839731079475115992,
\nonumber	\\
	\mathcal{C}_{a,i}^{FFA}(x_0) &= -93.20383029863213071037614923448,
\nonumber		\\
	\mathcal{C}_{t,ij}^{FFA}(x_0) &= -100.17493721704132096830414054544,
	\nonumber	\\
	\mathcal{C}_{t5,i0}^{FFA}(x_0) &= -100.05825656056622010641036389629,
\nonumber	\\
	\mathcal{C}_{a,i}^{FAA}(x_0) &= -63.723307019645252418839033187656,
	\nonumber	\\
	\mathcal{C}_{t,ij}^{FAA}(x_0) =
	\mathcal{C}_{t5,i0}^{FAA}(x_0) &= -60.51214557324242018544991891727,
	\nonumber	\\
	\mathcal{C}_{a,i}^{FFL}(x_0) &=30.264649695130721323661561264413 ,
	\nonumber	\\
	\mathcal{C}_{t,ij}^{FFL}(x_0) =
	\mathcal{C}_{t5,i0}^{FFL}(x_0) &=29.504022539029783506731432813772 ,
\nonumber	\\
	\mathcal{C}_{a,i}^{FFB}(x_0) &=0.46823995756914587629784660613504,
	\nonumber	\\
	\mathcal{C}_{t,ij}^{FFB}(x_0) =
	\mathcal{C}_{t5,i0}^{FFB}(x_0) &=0.27047996107038058906308237284441,
\nonumber	\\
	\mathcal{C}_{a,i}^{FFC}(x_0) &= 1.97214513321411357430121695665497 ,
	\nonumber	\\
	\mathcal{C}_{t,ij}^{FFC}(x_0) =
	\mathcal{C}_{t5,i0}^{FFC}(x_0) &= 1.28808262132866597109304979861138 ,
\nonumber	\\
	\mathcal{C}_{a,i}^{FAL}(x_0) &= 21.16033670918183535060836071863896,
	\nonumber	\\
	\mathcal{C}_{t,ij}^{FAL}(x_0) =
	\mathcal{C}_{t5,i0}^{FAL}(x_0) &= 20.9421929249525205645298703997716,
\nonumber	\\
	\mathcal{C}_{a,i}^{FAB}(x_0) &=-0.12883604932709405508825965377162 ,
	\nonumber	\\
	\mathcal{C}_{t,ij}^{FAB}(x_0) =
	\mathcal{C}_{t5,i0}^{FAB}(x_0) &= -0.112390888859631921955337734074188,
\nonumber	\\
	\mathcal{C}_{a,i}^{FAC}(x_0) &=-0.26483894578627489026464215544363 ,
	\nonumber	\\
	\mathcal{C}_{t,ij}^{FAC}(x_0) =
	\mathcal{C}_{t5,i0}^{FAC}(x_0) &=-0.19682239163578297497524199284322 ,
	\nonumber	\\
	\mathcal{C}_{a,i}^{FLL}(x_0) =	\mathcal{C}_{t,ij}^{FLL}(x_0) =	\mathcal{C}_{t5,i0}^{FLL}(x_0) &= -0.515699683439648770111279453569345,
\nonumber	\\
	\mathcal{C}_{a,i}^{FLB}(x_0) = 	\mathcal{C}_{t,ij}^{FLB}(x_0) =	\mathcal{C}_{t5,i0}^{FLB}(x_0) &=-0.0460089314692526820601917377234402,
\nonumber	\\
	\mathcal{C}_{a,i}^{FLC}(x_0) = 	\mathcal{C}_{t,ij}^{FLC}(x_0) =	\mathcal{C}_{t5,i0}^{FLC}(x_0) &=-0.315322209095270178269804879997075,
	\nonumber	\\
	\mathcal{C}_{a,i}^{FBB}(x_0) =	\mathcal{C}_{t,ij}^{FBB}(x_0) =	\mathcal{C}_{t5,i0}^{FBB}(x_0) &= -0.004731388929156892345278503644290267,
\nonumber	\\
	\mathcal{C}_{a,i}^{FBC}(x_0) =	\mathcal{C}_{t,ij}^{FBC}(x_0) =	\mathcal{C}_{t5,i0}^{FBC}(x_0) &= -0.0118687507006037939590736816982544,
\nonumber	\\
	\mathcal{C}_{a,i}^{FCC}(x_0) =	\mathcal{C}_{t,ij}^{FCC}(x_0) =	\mathcal{C}_{t5,i0}^{FCC}(x_0) &= 0.005739492344304311786485592038924699.
	\end{align}

From the above numerical values, one can see  the dominant contributions in $\mathcal{C}_J^{(2)}(x_0)$ and $\mathcal{C}_J^{(3)}(x_0)$ come from the components corresponding to the color factors  $C_F^2$, $C_F C_A$, $C_F^2C_A$ and $C_FC_A^2$, while the contributions from the bottom and charm quark loops are negligible~\cite{Tao:2022hos,Tao:2023mtw,Tao:2023vvf}.
We find almost all color-factor components between ${\cal C}_{t,ij}^{(n)}(x_0)$ and ${\cal C}_{t5,i0}^{(n)}(x_0)$ are exactly equal, except that ${\cal C}_{t,ij}^{FFF}(x_0)\approx{\cal C}_{t5,i0}^{FFF}(x_0)$ and ${\cal C}_{t,ij}^{FFA}(x_0)\approx{\cal C}_{t5,i0}^{FFA}(x_0)$.
Furthermore, we also observe that the color-factor  components corresponding to contributions from two-loop diagrams, each involving one quark loop, and three-loop diagrams, each involving two quark loops, are exactly equal among ${\cal C}_{a,i}^{(n)}(x_0)$, ${\cal C}_{t,ij}^{(n)}(x_0)$, and ${\cal C}_{t5,i0}^{(n)}(x_0)$, while the remaining color-factor  components of  ${\cal C}_{a,i}^{(n)}(x_0)$ are approximately equal to those of  ${\cal C}_{t,ij}^{(n)}(x_0)$ and ${\cal C}_{t5,i0}^{(n)}(x_0)$.

It's worth mentioning that in the calculations up to two loops, we  allow for a general QCD gauge parameter $\xi$   and have checked that   $\xi$ vanishes  in the final two-loop results of ${\overline {\cal C}}_J$ and ${\cal C}_{J}$ for $J\in\{(v,0),(t,ij),(t5,i0)\}$, which constitutes an important check on our calculations.
At the three-loop order, we work in Feynman gauge.
To further check our three-loop calculations, we have computed the three-loop matching coefficients at five rational numerical points~\cite{Feng:2022ruy,Sang:2022tnh,Tao:2022hos,Tao:2023mtw,Tao:2023vvf}: $x=\frac{150}{475}$, $\frac{475}{150}$, $\frac{204}{498}$, $\frac{498}{204}$, and $1$. 
With a precision of at least 30 significant digits,  our three-loop results at the five points of $x$ confirm that all of ${\overline {\cal C}}_J$, ${\cal C}_{J}$  and ${\cal C}_{J}^{(n)}(x)\,(n=1,2,3)$ for $J\in\{(v,0),(t,ij),(t5,i0)\}$  are invariant under the exchange  $m_b\leftrightarrow m_c$ meanwhile $n_b\leftrightarrow n_c$~\cite{Feng:2022ruy,Sang:2022tnh,Tao:2022hos,Tao:2023mtw,Tao:2023vvf}.

\section{Decay constant ratio~\label{Decayconstantratio}}

Within the NRQCD factorization formalism, the decay constant in QCD is factorized into the matching coefficient multiplied with the NRQCD long-distance matrix element (LDME), where the decay constant, as a physical quantity, should be perturbatively convergent while the matching coefficient and the LDME might be nonconvergent perturbative series~\cite{Tao:2023mtw,Beneke:2014qea,Rauh:2018vsv,Chung:2020zqc,Chung:2021efj,Chung:2023mgr,Sang:2023cwn}. 
%We consider the ratio of two decay constants and assume that 
%both the difference between the LDMEs and the difference between the masses of two distinct meson states within the same meson family are minor compared to  the difference between the  two corresponding matching coefficients and can be  neglected
We  consider the ratio of   two decay constants for  two  mesons coupled with two different currents within  the  same  meson  family, and assume that both  the  LDME  difference  and  the  mass  difference  between  the  two  mesons   are  negligible  compared  to  the matching coefficient difference  between  the  two  currents~\cite{Broadhurst:1994se,Bekavac:2009zc,Colquhoun:2015oha,Neubert:1992fk,Ball:1994uh,Campanario:2003ix,Tao:2023mtw,Collins:1999ff,Neubert:1993mb,Tao:2022qxa},    
%We consider the ratio of decay constants and  neglect the difference in the wave functions at the origin between two distinct states within the same meson family  as it is minor  in comparison to the difference in the  two corresponding matching coefficients~\cite{Broadhurst:1994se},  
which allows the unknown LDMEs  to be eliminated from the ratio~\cite{Beneke:1997jm,Onishchenko:2003ui}.
As a result,  the ratio of the physical decay constants is approximately equal to the ratio of the nonphysical matching coefficients~\cite{Broadhurst:1994se,Bekavac:2009zc,Colquhoun:2015oha,Ball:1994uh,Neubert:1993mb,Tao:2023vvf}.
So the ratio of two decay constants between different currents within the same meson family can be calculated by the following approximate formula:  
\begin{align}\label{ratiof}
\frac{f^{J_1}_{X_1}}{f^{J_2}_{X_2}}\approx\frac{\mathcal{C}_{J_1} \times \tilde{f}_{X_1}}{\mathcal{C}_{J_2} \times \tilde{f}_{X_2}}\approx\frac{\mathcal{C}_{J_1}}{\mathcal{C}_{J_2}},
\end{align}
where $f^{J}_{X}\left(\tilde{f}_{X}\right)$ is the decay constant (the LDME~\footnote{Note, the LDME  can be expressed in terms of the wave function at the origin and its derivatives. For the $S$-wave meson, $\tilde{f}_{S\text{-wave}}\sim|\Psi_{S\text{-wave}}(0)|$.  For the $P$-wave meson, $\tilde{f}_{P\text{-wave}}\sim|\nabla\Psi_{P\text{-wave}}(0)|$. See ref.~\cite{Chung:2020zqc,Chung:2021efj}  for more details.}) of the meson $X$ coupled with the current $J$ and the approximations are based on the assumption %~\cite{Broadhurst:1994se}:  
 $\tilde{f}_{X_1} \approx \tilde{f}_{X_2}$. 

With the matching coefficients for all ten heavy flavor-changing currents in eq.~\eqref{j10} known, we can study the ratios of $S$-wave and $P$-wave $c\bar b$ meson decay constants  involving scalar, pseudo-scalar, vector, axial-vector, tensor and axial-tensor currents.
In the following, 
we will present the phenomenological  results of the matching coefficients ${\cal C}_{v,0},{\cal C}_{t,ij},{\cal C}_{t5,i0}$, as well as the ratios of  matching coefficients (decay constants)  involving the ten currents coupled to the $S$-wave $B_c,B_c^*$ mesons and $P$-wave $B_{c0}^*,B_{c1}$ mesons.
%With the matching coefficients for all ten heavy flavor-changing currents known,
%we can study the ratios of decay constants for $S$-wave $B_c,B_c^*$ mesons and $P$-wave $B_{c0}^*,B_{c1}$ mesons involving various currents.
%In the following, 
%we will present the phenomenological  results of the matching coefficients ${\cal C}_{v,0},{\cal C}_{t,ij},{\cal C}_{t5,i0}$, as well as   the matching coefficient (decay constant) ratios for the $c\bar b$ mesons among the ten currents.
Throughout the remaining calculations, we expand the matching coefficients and the matching coefficient (decay constant) ratios in power series of $\alpha_s^{\left(n_l=3\right)}(\mu)$, and truncate them up to a fixed order~\cite{Tao:2023mtw,Tao:2023vvf}.

\begin{table}[htbp]
	\caption{The expansion coefficients of $\left({\alpha_s^{\left(n_l=3\right)}(\mu=\mu_0)}/{\pi}\right)^i~(i=0,1,2,3)$ with  $\mu_f=1.2\,\mathrm{GeV}$, $\mu=\mu_0=3\,\mathrm{GeV}$, $m_b=4.75\,\mathrm{GeV}$, $m_c=1.5\,\mathrm{GeV}$ for the matching coefficients  ${\cal C}_{v,0},{\cal C}_{t,ij},{\cal C}_{t5,i0}$, and    the matching coefficient (decay constant) ratios of $c \bar b$ mesons involving various currents. 
	 	Note that    	${f_{B_c^*}^{v,i}}/{f_{B_{c0}^*}^{s}}\not\approx{\mathcal{C}_{v,i}}/{\mathcal{C}_{s}} $.    }
	\label{tab:asexpandnum}
	\setlength{\tabcolsep}{1.0mm}
	\centering
	\renewcommand{\arraystretch}{1.2}%{2.0}
	%\resizebox{\textwidth}{!}
	{
		\begin{tabular}{c|c|c|c|c}
			\hline%\hline
			{$$}
			& $\left({\alpha_s^{\left(3\right)}(\mu_0)}/{\pi}\right)^0$
			& $\left({\alpha_s^{\left(3\right)}(\mu_0)}/{\pi}\right)^1$
			& $\left({\alpha_s^{\left(3\right)}(\mu_0)}/{\pi}\right)^2$ 
			& $\left({\alpha_s^{\left(3\right)}(\mu_0)}/{\pi}\right)^3$   \\
			\hline
			%----------------------------------------------------------
			\multirow{1}*{$	\mathcal{C}_{v,0}$}
			& \multirow{1}*{$1$}
			& \multirow{1}*{$1.550025$}
			& \multirow{1}*{$- 	4.432751$}
			& \multirow{1}*{$- 						841.9545$}               \\
			%			&
			%			&
			%			& 
			%			&   \\
			\hline
			%----------------------------------------------------------
			{$	\mathcal{C}_{t,ij} $}
			& {$1$}
			& {$- 0.7339400$}
			&{$ - 				18.87492$}
			& {      $ - 		914.8718$     }              \\
			%			&
			%			&
			%			& 
			%			& \\
			\hline
			%--------------------------------------------------------
			{$\mathcal{C}_{t5,i0}$}
			& {$1$}
			& {$- 0.7339400 $}
			&{$- 						18.87492 $}
			&   {    $- 					914.1073 $  }                 \\
			%			&
			%			&
			%			& 
			%			& \\
			\hline
			%--------------------------------------------------------
			%{$\frac{f_{B_c^*}^{v,i}}{f_{B_{c0}^*}^{s}}\approx\frac{\mathcal{C}_{v,i}}{\mathcal{C}_{s}}$}
			${\mathcal{C}_{v,i}}/{\mathcal{C}_{s}}$
			& {$1$}
			& {$-2$}
			&{$ - 	13.96230$}
			&   {    $  - 		786.2656$  }                 \\
			%			&
			%			&
			%			& 
			%			& \\
			\hline
			%--------------------------------------------------------
			{$	{f_{B_c}^p}/{f_{B_c^*}^{v,i}}\approx{\mathcal{C}_{p}}/{\mathcal{C}_{v,i}} $}
			& {$1$}
			& {$ 0.6666667$}
			&{$	2.869082 $}
			&   {    $ - 	65.95654$  }                 \\
			%			&
			%			&
			%			& 
			%			& \\
			\hline
			%--------------------------------------------------------
			{${f_{B_{c0}^*}^{s}}/{f_{B_{c1}}^{a,i}}\approx{\mathcal{C}_{s}}/{\mathcal{C}_{a,i}}$}
			& {$1$}
			& {$ 0.6666667$}
			&{$	3.193040$}
			&   {    $	1.731840$  }                 \\
			%			&
			%			&
			%			& 
			%			& \\
			\hline
			%--------------------------------------------------------
			{${f_{B_{c0}^*}^{s}}/{f_{B_{c0}^*}^{v,0}}\approx{\mathcal{C}_{s}}/{\mathcal{C}_{v,0}}$}
			& {$1$}
			& {$- 1.617298$}
			&{$- 	8.524304 $}
			&   {    $- 	87.47028$  }                 \\
			%			&
			%			&
			%			& 
			%			& \\
			\hline
			%--------------------------------------------------------
			{${f_{B_{c}^*}^{t,i0}}/{f_{B_{c}^*}^{t5,ij}}\approx{\mathcal{C}_{t,i0}}/{\mathcal{C}_{t5,ij}}$}
			& {$1$}
			& {$0$}
			&{$	2.133589$}
			&   {    $  		11.04305 $  }                 \\
			%			&
			%			&
			%			& 
			%			& \\
			\hline
			%--------------------------------------------------------
			{${f_{B_{c1}}^{a,i}}/{f_{B_{c1}}^{t5,i0}}\approx{\mathcal{C}_{a,i}}/{\mathcal{C}_{t5,i0}}$}
			& {$1$}
			& {$0$}
			&{$	0.7072639$}
			&   {    $ - 		8.118842 $  }                 \\
			%			&
			%			&
			%			& 
			%			& \\
			\hline
			%--------------------------------------------------------
			{${f_{B_{c1}}^{t,ij}}/{f_{B_{c1}}^{t5,i0}}\approx{\mathcal{C}_{t,ij}}/{\mathcal{C}_{t5,i0}}$}
			& {$1$}
			& {$0$}
			&{$0$}
			&   {    $ - 0.7645181$  }                 \\
			%			&
			%			&
			%			& 
			%			& \\
			\hline%\hline
			\end{tabular}
		}
		\end{table}

\begin{figure}[htbp]
	\centering
	\includegraphics[width=0.44\textwidth]{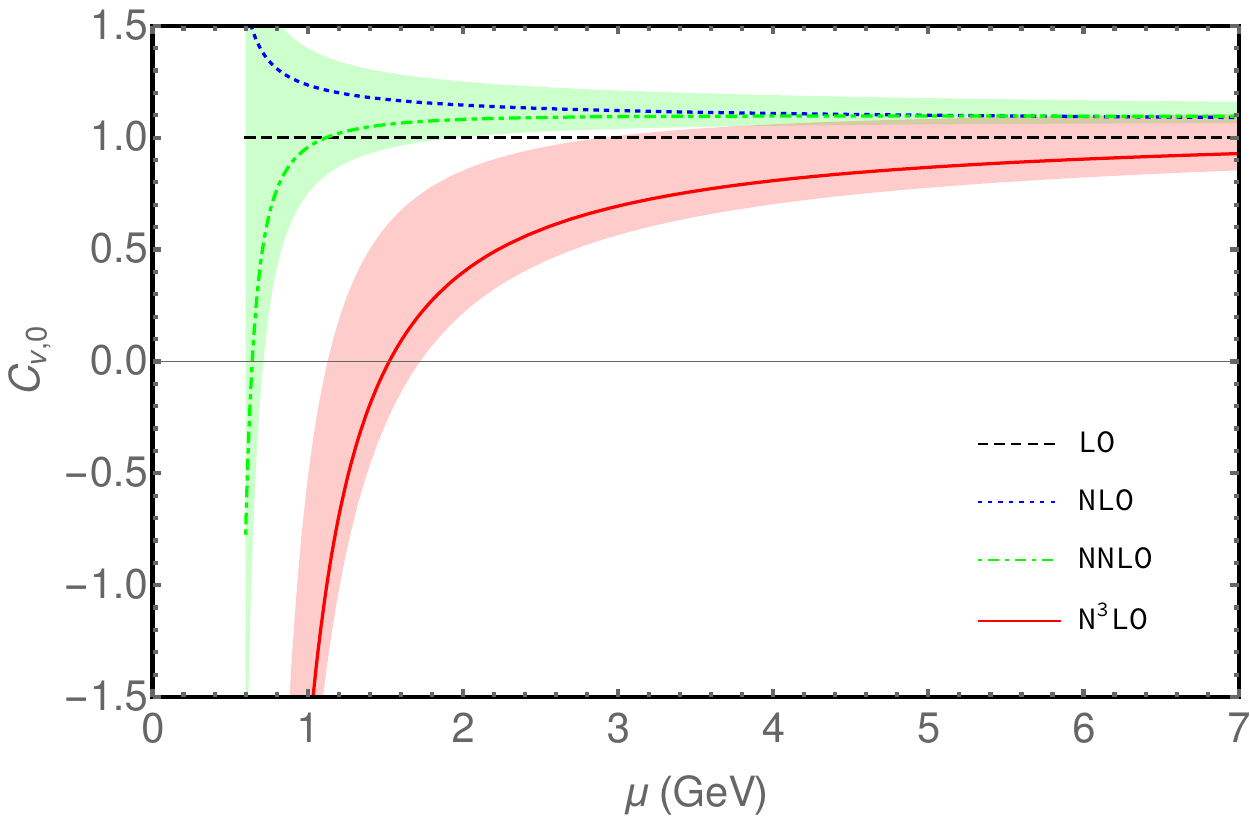}	\\
	\includegraphics[width=0.44\textwidth]{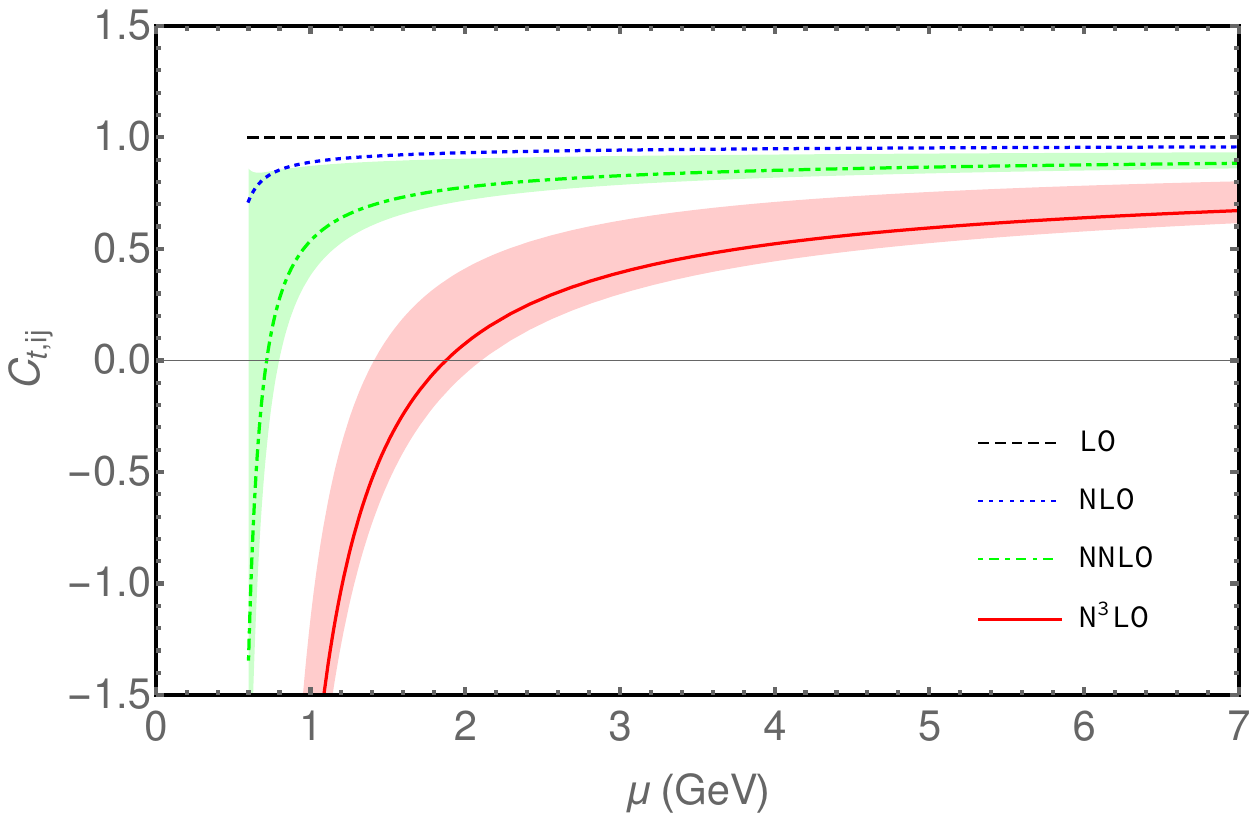}\qquad
	\includegraphics[width=0.44\textwidth]{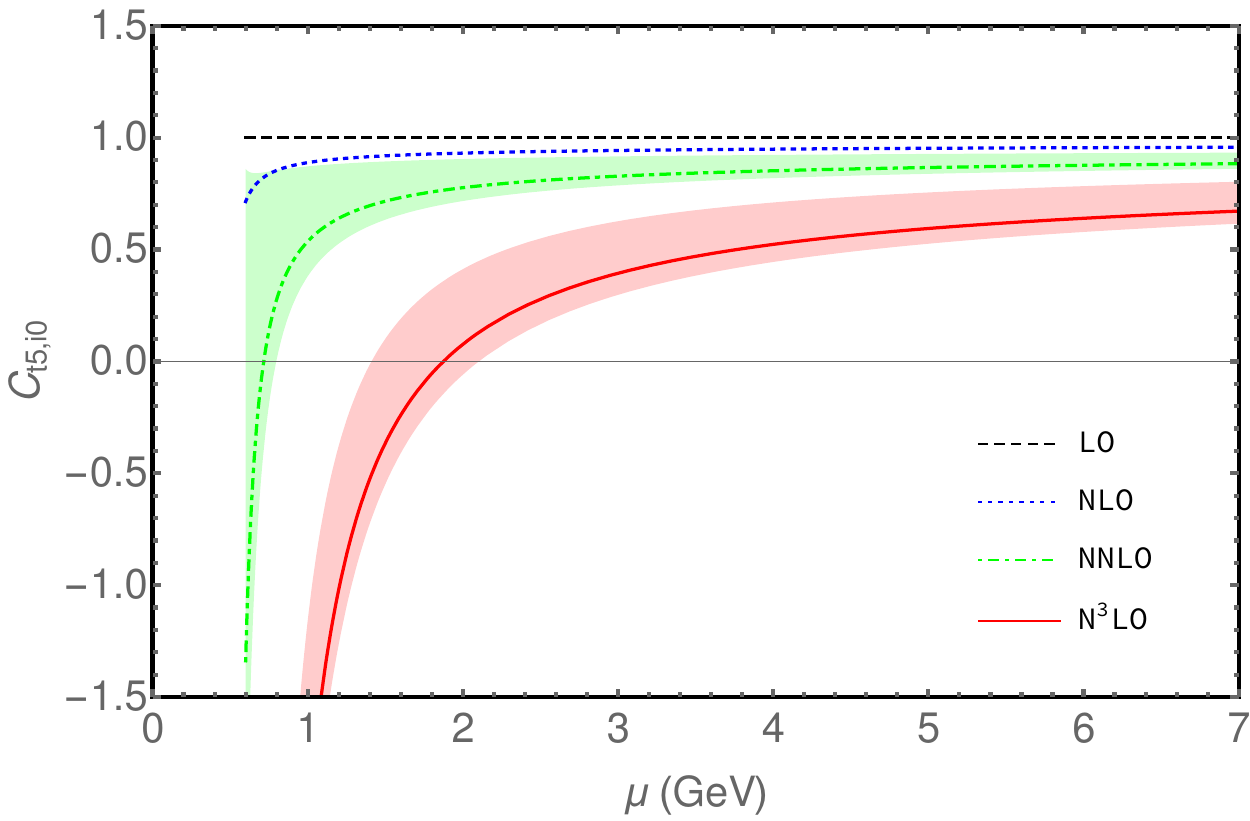}\qquad
	\caption{The QCD renormalization scale $\mu$ dependence of the matching coefficients $\mathcal{C}_{J}$ for $J\in\{(v,0),(t,ij),(t5,i0)\}$
		at LO,  NLO,  NNLO and N$^3$LO accuracy. 
		The central values of  the matching coefficients  are calculated with the  physical values:  $\mu_f=1.2\,\mathrm{GeV}$,  $m_b=4.75\,\mathrm{GeV}$ and $m_c=1.5\,\mathrm{GeV}$. The error bands stem from varying  the NRQCD factorization scale $\mu_f$ between    0.4  and 2    $\mathrm{GeV}$, corresponding to      the bands' upper and lower  edges, respectively.  
%The error bands stem from the variation of  the NRQCD factorization scale $\mu_f$,
%with the upper and lower  edges	of  the  bands  corresponding to         
%$\mu_f^{min}=0.4\,\mathrm{GeV}$ and $\mu_f^{max}=2\,\mathrm{GeV}$, respectively.	
}
	\label{fig:Cjmu}
\end{figure}

\begin{figure}[htbp]%[H]
	\centering
	\includegraphics[width=0.44\textwidth]{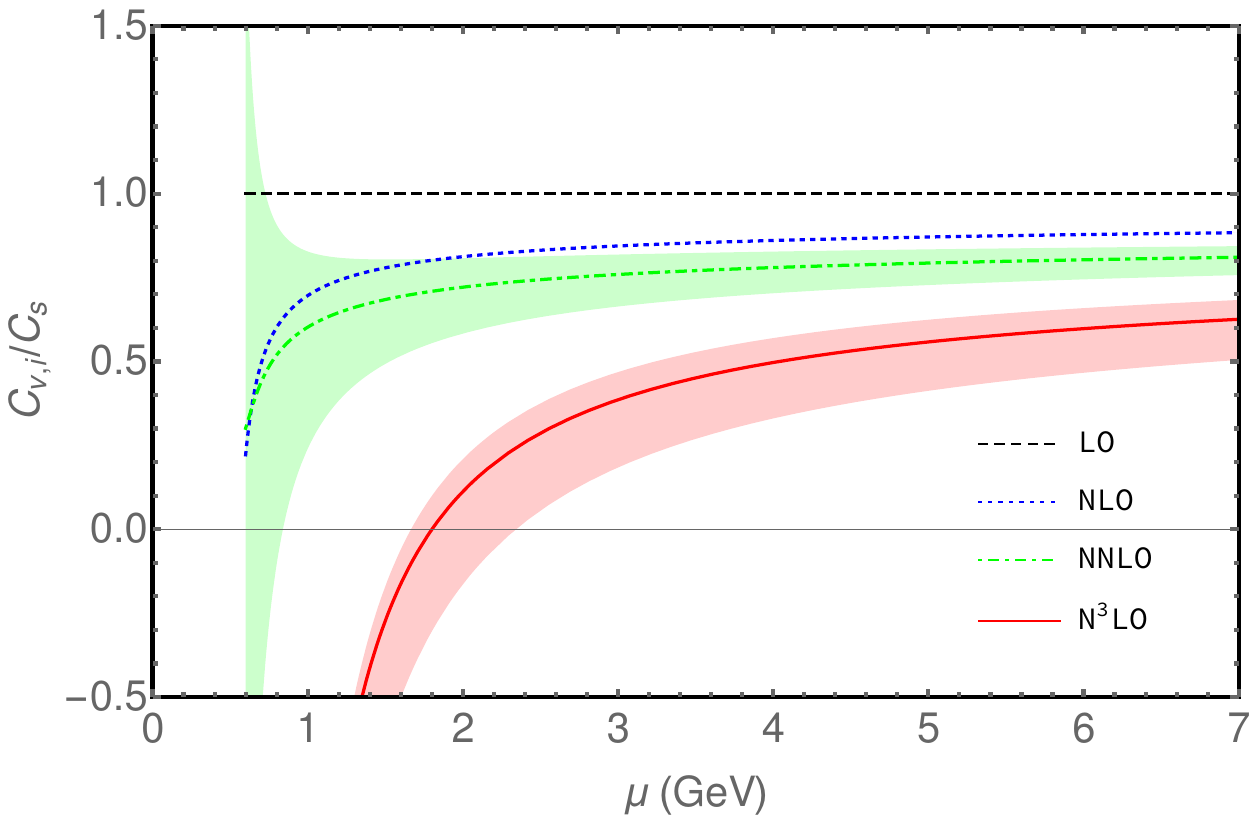}\\
	\includegraphics[width=0.44\textwidth]{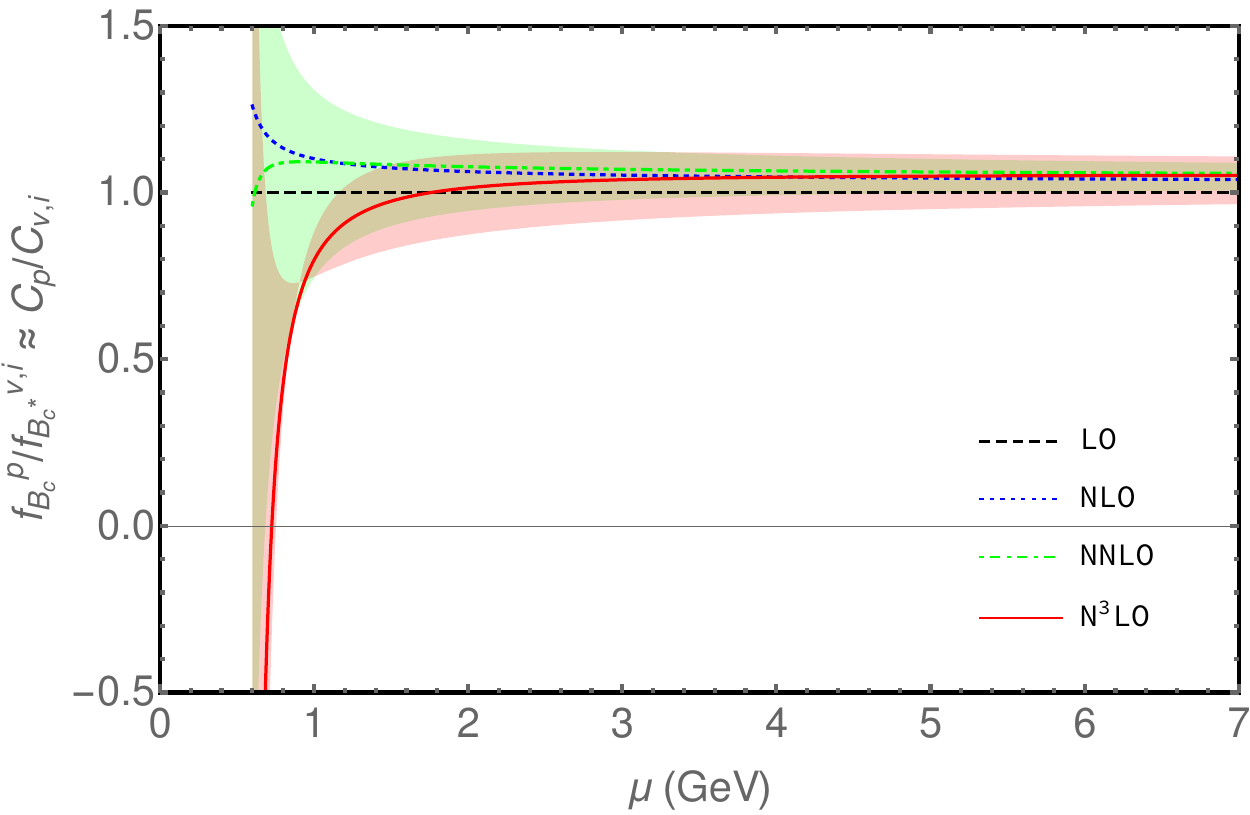}\qquad
	\includegraphics[width=0.44\textwidth]{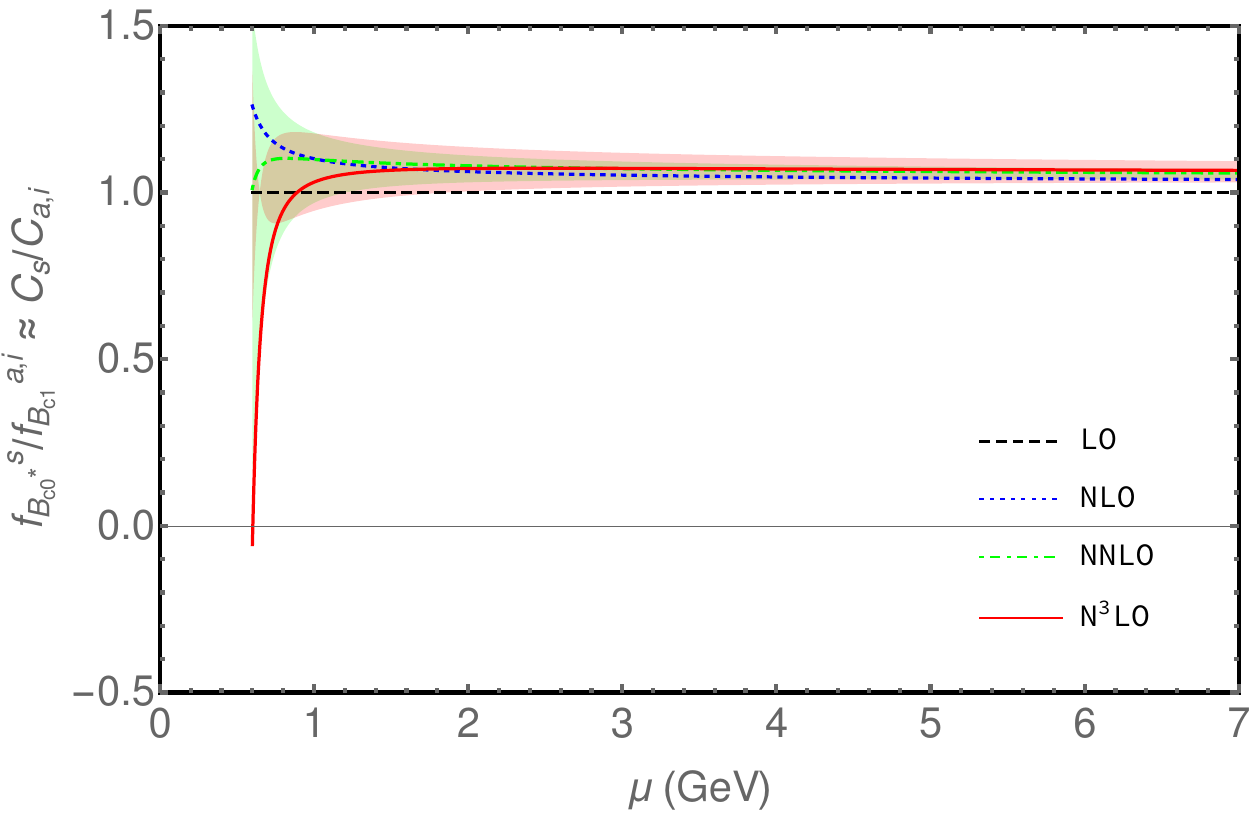}\qquad
	\includegraphics[width=0.44\textwidth]{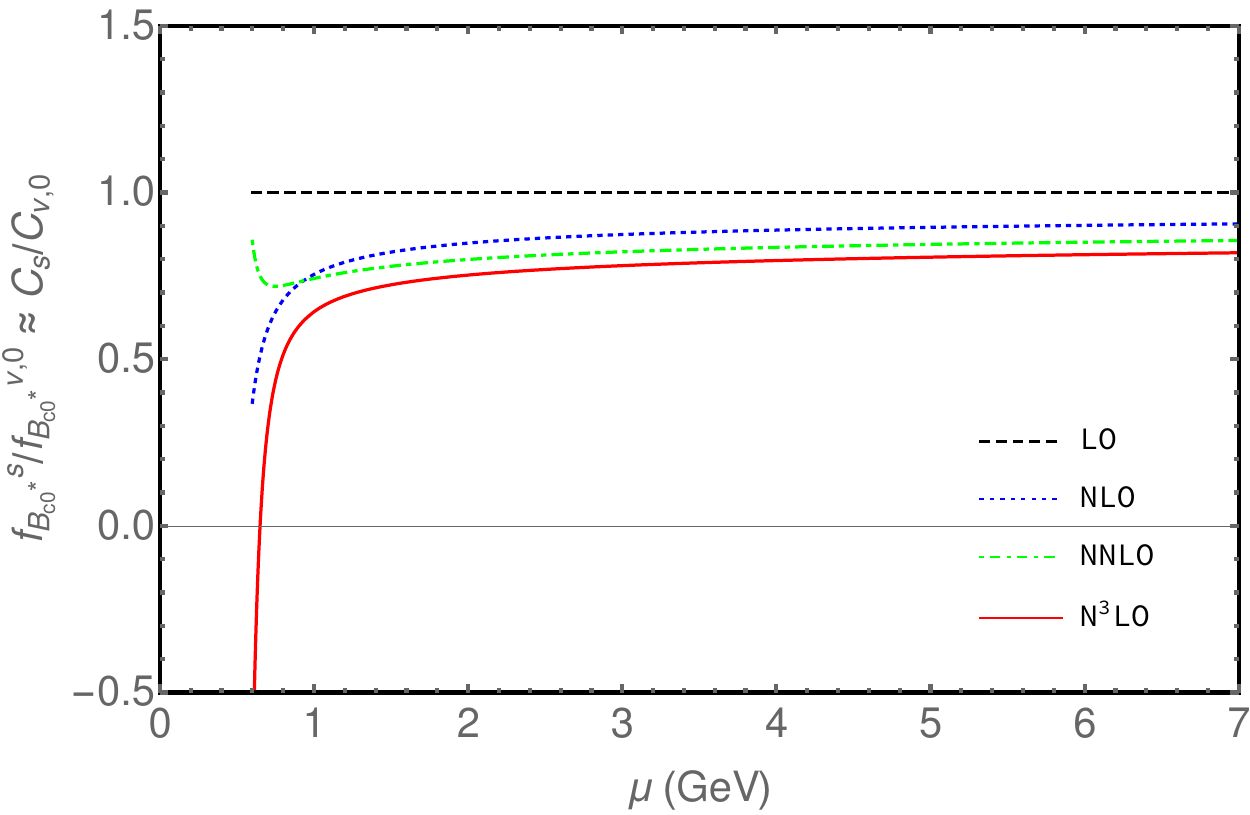}\qquad
	\includegraphics[width=0.44\textwidth]{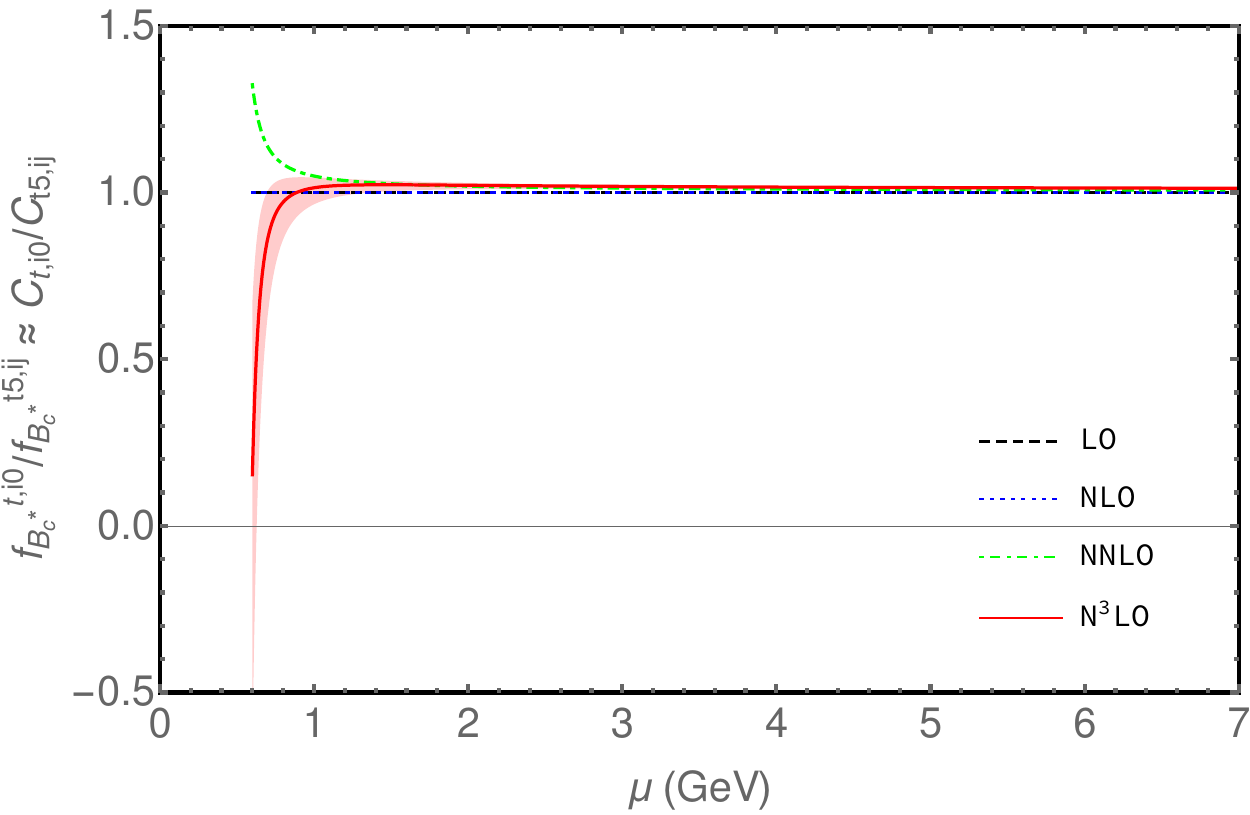}\qquad	
	\includegraphics[width=0.44\textwidth]{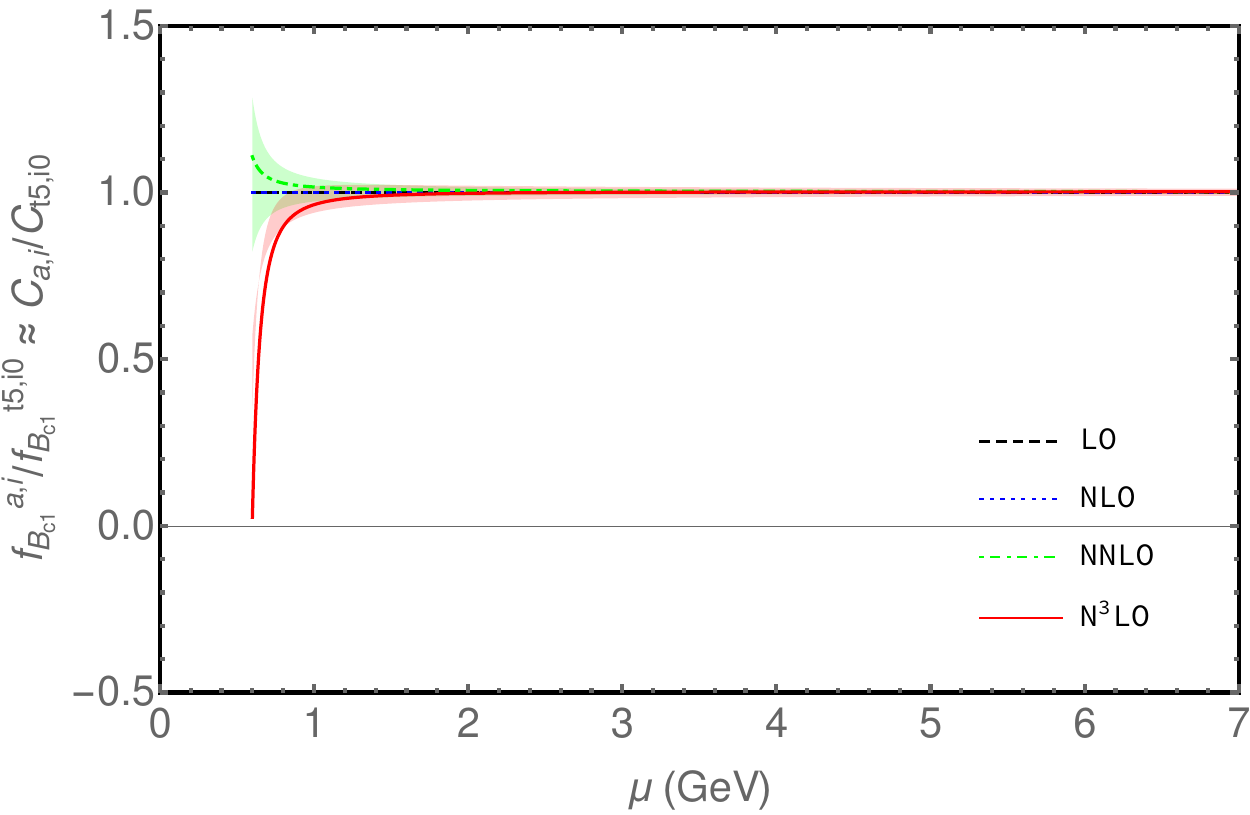}\qquad
	\includegraphics[width=0.44\textwidth]{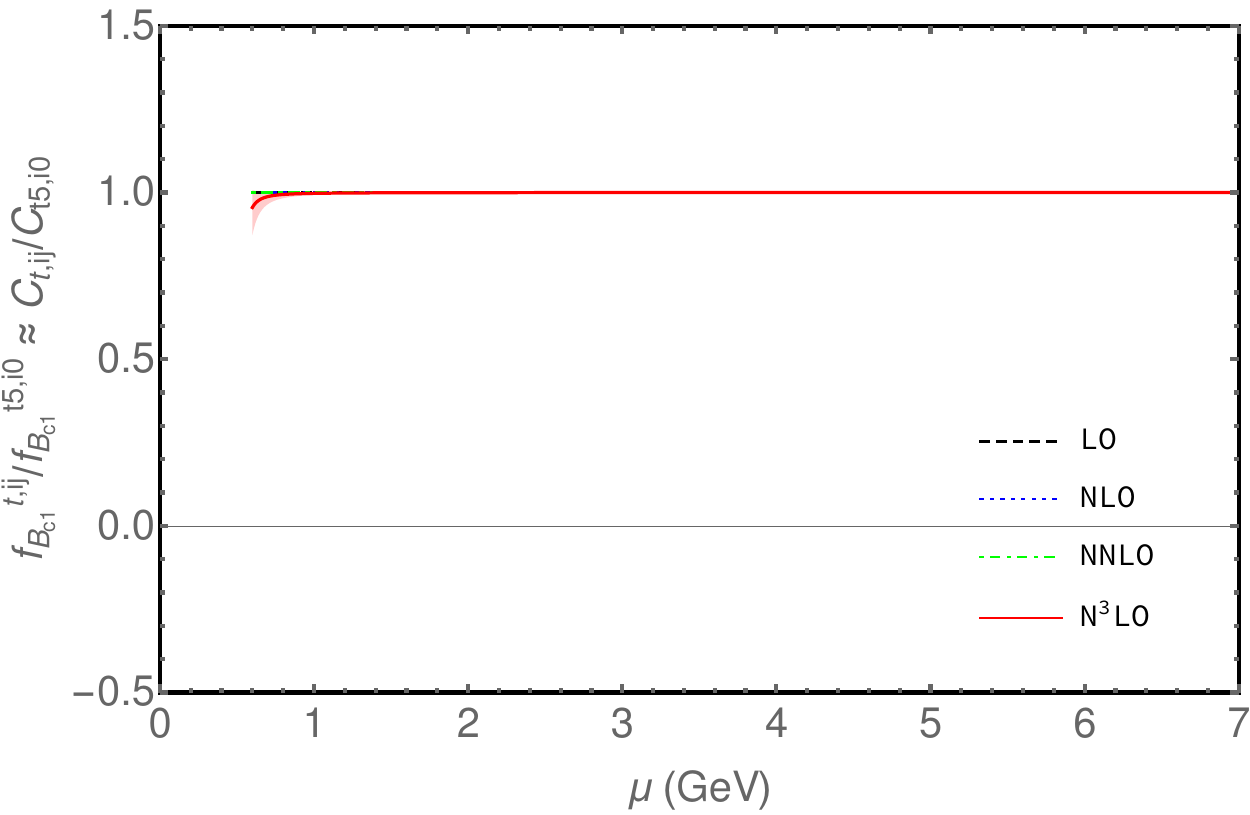}\qquad
	\caption{The QCD renormalization scale $\mu$ dependence of the matching coefficient (decay constant) ratios for $c\bar b$ mesons involving various  currents  
		at LO,  NLO,  NNLO and N$^3$LO accuracy. The central values of the matching coefficient (decay constant) ratios are calculated with the  physical values:   $\mu_f=1.2\,\mathrm{GeV}$,  $m_b=4.75\,\mathrm{GeV}$ and $m_c=1.5\,\mathrm{GeV}$.   
		The error bands stem from varying   $\mu_f$ between    0.4  and 7    $\mathrm{GeV}$, corresponding to      the bands' upper and lower  edges, respectively. 
	 	Note that    	$\frac{f_{B_c^*}^{v,i}}{f_{B_{c0}^*}^{s}}\not\approx\frac{\mathcal{C}_{v,i}}{\mathcal{C}_{s}} $. }
	\label{fig:CivCjmu}
\end{figure}

\begin{table}[htbp]%[H]
	\begin{center}
		\caption{The perturbative QCD predictions of  the matching coefficient (decay constant)  ratios for $c\bar b$ mesons involving various currents at LO, NLO, NNLO and N$^3$LO accuracy.
			The central values of  the matching coefficient (decay constant)  ratios  are calculated %using the  physical values with
			at the physical point $\mu_f=1.2\,\mathrm{GeV}$, $\mu=\mu_0=3\,\mathrm{GeV}$, $m_b=4.75\,\mathrm{GeV}$ and $m_c=1.5\,\mathrm{GeV}$.
			The uncertainties in the first and second columns are  estimated by varying $\mu_f$  from   7 to 0.4 $\mathrm{GeV}$ and $\mu$  from   7 to 1.5 $\mathrm{GeV}$,   respectively.
			Note that    	$\frac{f_{B_c^*}^{v,i}}{f_{B_{c0}^*}^{s}}\not\approx\frac{\mathcal{C}_{v,i}}{\mathcal{C}_{s}} $.	  }
		\label{tab:CivCjnum}
		\renewcommand\arraystretch{1.6}
		\tabcolsep=1.8mm
		\begin{tabular}{ c c c c c}
			\hline%\hline
			& LO         &  NLO                   & NNLO     & N$^3$LO	
			\\  \hline
			%$\frac{f_{B_c^*}^{v,i}}{f_{B_{c0}^*}^{s}}\approx\frac{\mathcal{C}_{v,i}}{\mathcal{C}_{s}} $	
			$\frac{\mathcal{C}_{v,i}}{\mathcal{C}_{s}} $
			& $1$ & $0.84400^{-0+0.03953}_{+0-0.06519}$   &    $0.75905^{-0.09541+0.05127}_{+0.05944-0.07471}$  &     $0.38591^{-0.20176+0.23967}_{+0.08251-0.65866}$
			\\  \hline
			$\frac{f_{B_c}^p}{f_{B_c^*}^{v,i}}\approx\frac{\mathcal{C}_{p}}{\mathcal{C}_{v,i}}$	 & $1$ & $1.05200^{-0-0.01318}_{+0+0.02173}$   &    $1.06946^{-0.09157-0.01228}_{+0.05704+0.01393}$  &     $1.03816^{-0.12195+0.01312}_{+0.08313-0.06443}$
			\\  \hline
			$\frac{f_{B_{c0}^*}^{s}}{f_{B_{c1}}^{a,i}}\approx\frac{\mathcal{C}_{s}}{\mathcal{C}_{a,i}} $	 & $1$ & $1.05200^{-0-0.01318}_{+0+0.02173}$   &    $1.07143^{-0.03434-0.01315}_{+0.02139+0.01592}$  &     $1.07225^{-0.05482-0.00602}_{+0.04668+0}$
			\\  \hline
			$\frac{f_{B_{c0}^*}^{s}}{f_{B_{c0}^*}^{v,0}}\approx\frac{\mathcal{C}_{s}}{\mathcal{C}_{v,0}}$	 & $1$ & $0.87385^{-0+0.03197}_{+0-0.05271}$   &    $0.82199^{-0+0.03401}_{+0-0.04341}$  &     $0.78047^{-0+0.03860}_{+0-0.05744}$
			\\  \hline
			$\frac{f_{B_{c}^*}^{t,i0}}{f_{B_{c}^*}^{t5,ij}}\approx\frac{\mathcal{C}_{t,i0}}{\mathcal{C}_{t5,ij}}$	 & $1$ & $1$   &    $1.01298^{-0-0.00575}_{+0+0.01312}$  &     $1.01822^{-0.00670-0.00559}_{+0.00417+0.00481}$
			\\  \hline
			$\frac{f_{B_{c1}}^{a,i}}{f_{B_{c1}}^{t5,i0}}\approx\frac{\mathcal{C}_{a,i}}{\mathcal{C}_{t5,i0}}$	 & $1$ & $1$   &    $1.00430^{-0.01145-0.00190}_{+0.00713+0.00435}$  &     $1.00045^{-0.01767+0.00141}_{+0.01638-0.00875}$
			\\  \hline
			$\frac{f_{B_{c1}}^{t,ij}}{f_{B_{c1}}^{t5,i0}}\approx\frac{\mathcal{C}_{t,ij}}{\mathcal{C}_{t5,i0}}$	 & $1$ & $1$   &    $1$  &     $0.99964^{-0.00067+0.00021}_{+0.00042-0.00067}$
			\\		\hline %\hline
		\end{tabular}
	\end{center}
\end{table}

In table~\ref{tab:asexpandnum},  we present the   $\alpha_s$-expansions of both the matching coefficients    and   the matching coefficient (decay constant) ratios  at the physical point: $\mu_f=1.2\,\mathrm{GeV}$, $\mu=\mu_0=3\,\mathrm{GeV}$, $m_b=4.75\,\mathrm{GeV}$, and $m_c=1.5\,\mathrm{GeV}$.  
By numerically evaluating $\alpha_s^{\left(n_l=3\right)}(\mu)$ with the aid of the \texttt{RunDec} package, 
%With $\alpha_s^{\left(n_l=3\right)}(\mu)$ numerically evaluated via the \texttt{RunDec} package, 
as detailed in Sec.~\ref{ZjNRQCD}, we proceed to investigate the dependence on the QCD renormalization scale $\mu$. We present the $\mu$ dependence at LO, NLO, NNLO, and N$^3$LO accuracy for the matching coefficients in figure~\ref{fig:Cjmu}, as well as the matching coefficient (decay constant) ratios in figure~\ref{fig:CivCjmu}. Finally, we provide the perturbative QCD predictions up to N$^3$LO for the matching coefficient (decay constant) ratios  of the $c\bar b$ mesons involving various heavy flavor-changing currents in table~\ref{tab:CivCjnum}.

Based  on tables~\ref{tab:asexpandnum} and~\ref{tab:CivCjnum}, as well as figures~\ref{fig:Cjmu} and~\ref{fig:CivCjmu}, along with the corresponding results of matching coefficients for the remaining seven currents as  presented in our earlier works~\cite{Tao:2023mtw,Tao:2023vvf}, 
%(eq. (5.13) and figure 3 in ref.~\cite{Tao:2023mtw}, eq. (28) and figure 5 in ref.~\cite{Tao:2023vvf}), 
we make the following key observations and insights:
\begin{enumerate}
	\item[(1)]

The perturbative expansion of the matching coefficient for each of the ten currents does not converge up to N$^3$LO. Specifically, the correction of order $\alpha_s^3$ is very large.
Moreover, the N$^3$LO correction for each of the ten currents exhibits  very strong dependence on  the QCD renormalization scale $\mu$, as well as the NRQCD factorization scale $\mu_f$. 

\item[(2)]

As shown in table~\ref{tab:asexpandnum}, figure~\ref{fig:CivCjmu} and table~\ref{tab:CivCjnum}, 
%Figure~\ref{fig:CivCjmu} and table~\ref{tab:CivCjnum}  demonstrate     that
for the seven ratios of matching coefficients among the ten currents,  excluding ${\cal C}_{v,i}/{\cal C}_{s}$, the N$^3$LO correction to each of the remaining six ratios is convergent and has very weak dependence on both $\mu$ and $\mu_f$.
% are convergent  up to ${\mathcal O}(\alpha_s^3)$
Given that the physical decay constant ratio is convergent and scale-independent, 
the good perturbative convergence and weak scale-dependence of the matching coefficient ratio suggest that the difference between the corresponding  two LDMEs is relatively small and can be neglected~\cite{Tao:2023mtw}.  Therefore,    eq.~\eqref{ratiof}  holds true in  this case,  i.e., the six matching coefficient ratios are approximately equal  to their corresponding  decay constant ratios.   
%As a result,  the six decay constant ratios, each converging up to N$^3$LO, can be obtained.
%As a result, the obtained six decay constant  ratios, each converging up to N$^3$LO, are reliable.
Thus, we can obtain the  N$^3$LO result, which has good convergence behavior with weak  scale-dependence, for each of the six decay constant ratios.  The  reliability of our results  can be further confirmed by cross checks. 
For example,  
the N$^3$LO result of   ${f_{B_c}^{p}}/{f_{B_c^*}^{v,i}}$ in table~\ref{tab:CivCjnum} is in good agreement  with the latest Lattice result
 ${f_{B_c}^{p}}/{f_{B_c^*}^{v,i}}={1}/{0.988(27)}$~\cite{Colquhoun:2015oha}, as well as our previous calculation~\cite{Tao:2023mtw}.
Additionally, the results in figure~\ref{fig:CivCjmu} and table~\ref{tab:CivCjnum}  show that ${f_{B_{c0}^*}^{s}}>{f_{B_{c1}}^{a,i}}\gtrsim f_{B_{c1}}^{t5,i0}$, which is consistent with relevant literature~\cite{Verma:2011yw,Chung:2021efj,Hwang:2012nw,Sundu:2011vz,Pullin:2021ebn,Yang:2007zt,Olpak:2016wkf,Hwang:2009cu}.

\item[(3)]

From figure~\ref{fig:CivCjmu} and table~\ref{tab:CivCjnum},
%the matching coefficient ratio between spatial vector and scalar currents
%the ratio of matching coefficients between spatial vector $(v,i)$ and scalar $(s)$ currents 
%the ratio of matching coefficients for spatial vector to scalar currents   
the spatial vector $(v,i)$ to scalar $(s)$ current matching coefficient ratio ${\cal C}_{v,i}/{\cal C}_s$ 
%the ratio of spatial vector $(v,i)$ to scalar (s) current matching coefficients
%the matching coefficient ratio of spatial vector to scalar currents 
%$({\cal C}_{v,i}/{\cal C}_s)$ 
is still nonconvergent up to ${\mathcal O}(\alpha_s^3)$ and also exhibits strong dependence on both $\mu$ and $\mu_f$, 
which implies the difference between the LDME of the $S$-wave $B_c^*$ meson and that of the $P$-wave $B_{c0}^*$ meson is rather large and can not be neglected, i.e. $\tilde{f}_{B_c^*}\not\approx\tilde{f}_{B_{c0}^*}$.
Therefore, eq.~\eqref{ratiof} does not hold for the ratio of $(v,i)$ to $(s)$, i.e.
${f_{B_c^*}^{v,i}}/{f_{B_{c0}^*}^{s}}\not\approx{\mathcal{C}_{v,i}}/{\mathcal{C}_{s}} $.
Furthermore, figure~\ref{fig:CivCjmu}  displays ${\mathcal{C}_{v,i}}/{\mathcal{C}_{s}}<1$, while a substantial body of literature consistently reports ${f_{B_c^*}^{v,i}}/{f_{B_{c0}^*}^{s}}>1$~\cite{Chung:2020zqc,Chung:2021efj,Wang:2005qx,Wang:2007av,Wang:2008as,Verma:2011yw,Aliev:1992vp,Wang:2012kw,Narison:2020wql,Chang:2018zjq}.  
Thus, the correct N$^3$LO result of ${f_{B_c^*}^{v,i}}/{f_{B_{c0}^*}^{s}}$ cannot be  obtained from ${\mathcal{C}_{v,i}}/{\mathcal{C}_{s}}$ and eq.~\eqref{ratiof}.   
However,    considering   the meson mass hierarchy ($m_{B_c} < m_{B_c^*} < m_{B_{c0}^*} < m_{B_{c1}}$~\cite{Dowdall:2012ab,Martin-Gonzalez:2022qwd,Wang:2012kw})  and the scale relation (in the $c\bar{b}$ family, the greater the meson mass, the smaller the decay constant~\cite{Kiselev:1995bv}), we ultimately conclude that  the decay constant of the $S$-wave $c\bar{b}$ meson is greater than that of the $P$-wave, i.e. $f_{S\text{-wave}}>f_{P\text{-wave}}$, as evidenced by the specific example of  ${f_{B_c^*}^{v,i}}>{f_{B_{c0}^*}^{s}}$ and supported by related studies in 
various literature~\cite{Chung:2020zqc,Chung:2021efj,Becirevic:2013bsa,Wang:2005qx,Wang:2007av,Wang:2008as,Verma:2011yw,Aliev:1992vp,Wang:2012kw,Narison:2020wql,Chang:2018zjq,Agaev:2017lmc}.

\item[(4)]

%As shown in figure 3, a, b, c, and  the remaining four matching coefficient ratios  have   strong, weak,  weaker, and almost invisible   dependence   on both uf and u, along with  poor,  good,  better, and  nearly optimal  convergence, respectively. 
%As shown in Figure 1, from cvi/cs  to cp/cvi  to  cs/cai, and to the remaining four matching coefficient ratios,   the dependence on both uf and u is reduced progressively  and the convergence is improved gradually. 

%As shown in figure~\ref{fig:CivCjmu}, the dependence of  ${\cal C}_{v,i}/{\cal C}_{s}$, ${\cal C}_p/{\cal C}_{v,i}$, ${\cal C}_s/{\cal C}_{a,i}$, and the remaining four matching coefficient ratios on both $\mu_f$ and $\mu$ ranges from strong, to weak, to  weaker, and to almost invisible, respectively. Meanwhile, their convergence also varies from poor, to good, to better, and to nearly optimal, respectively. 
%From the variations of the scale dependence and convergence among the seven matching coefficient ratios, 
%and considering that both the physical decay constant and decay constant ratio are convergent and scale-independent, 
%we deduce that
Figure~\ref{fig:CivCjmu}   also illustrates the variations of the convergence and scale dependence among the seven matching coefficient ratios. Notably, ${{\cal C}_{v,i}}/{{\cal C}_s}$ stands out with the poorest convergence and the strongest dependence on both $\mu_f$ and $\mu$. In contrast,  ${{\cal C}_s}/{{\cal C}_{a,i}}$ demonstrates comparatively  better convergence and relatively weaker dependence on both $\mu_f$ and $\mu$ than ${{\cal C}_p}/{{\cal C}_{v,i}}$.  Additionally, the remaining four matching coefficient ratios display nearly  optimal convergence,  
%and are almost independent of both $\mu_f$ and $\mu$.  
and their dependence on both $\mu_f$ and $\mu$ is almost invisible, i.e. they are almost scale-independent.
From these variations, and considering that both the physical decay constant and decay constant ratio are convergent and scale-independent,    we deduce that 
%From figure~\ref{fig:CivCjmu}, we also find
for the $c \bar b$ meson states with different orbital/spin/total angular momentum quantum numbers, the LDMEs have large/small/smaller differences,
i.e., 
the LDME  difference  between $S$-wave and $P$-wave states is significantly greater than that  within $S$-wave states, which in turn  is slightly greater than that within $P$-wave states. Symbolically speaking, 
$\Delta(\tilde{f}_{B_c(B_c^*)}, \tilde{f}_{B_{c0}^*(B_{c1})})\gg\Delta(\tilde{f}_{B_c}, \tilde{f}_{B_c^*})>\Delta(\tilde{f}_{B_{c0}^*},\tilde{f}_{B_{c1}})$, 
where  $\Delta(x_1,x_2)$ denotes the difference between  $x_1$ and $x_2$.  
$\tilde{f}_{B_{c}},\tilde{f}_{B_{c}^*},\tilde{f}_{B_{c0}^*}$ and $\tilde{f}_{B_{c1}}$ are the LDMEs for  $B_c(^1S_0),B^*_c(^3{S_1}),B^*_{c0}({^3P_0})$ and $B_{c1}(^3P_1)$, respectively. 
For the same $c \bar b$ meson state coupled with different currents, the LDME remains the same. 
Therefore, by eq.~\eqref{ratiof}, 
%the credibility of the obtained N$^3$LO results for the six  decay constant ratios improves progressively 
the credibility of our N$^3$LO results  improves progressively
from $f_{B_c}^p/f_{B_c^*}^{v,i}$ to $f_{B_{c0}^*}^s/f_{B_{c1}}^{a,i}$, and then to the remaining four obtained decay constant ratios ($f_{B_{c0}^*}^s/f_{B_{c0}^*}^{v,0},f_{B_c^*}^{t,i0}/f_{B_c^*}^{t5,ij},f_{B_{c1}}^{a,i}/f_{B_{c1}}^{t5,i0}$, and $f_{B_{c1}}^{t,ij}/f_{B_{c1}}^{t5,i0}$).

%Substantial, minor, and minimal differences are observed in the wave functions at the origin among bc meson states with varying orbital, spin, and total angular momentum quantum numbers. Additionally, the wave function at the origin remains consistent for the same bc meson state coupled with different currents.

\item[(5)]

Combining the relations of decay constants in eq.~\eqref{NRQCDrelations} and our N$^3$LO results of the six decay constant ratios  in figure~\ref{fig:CivCjmu} and table~\ref{tab:CivCjnum}  with our conclusion $f_{S\text{-wave}}>f_{P\text{-wave}}$, 
we provide the NRQCD prediction for the hierarchical relationship among 
 % the ordering of values for ...
%we predict the  relationships  among 
the decay constants of the $c\bar b$ mesons involving the ten heavy flavor-changing currents: 
%as follows:
\begin{align}  
f_{B_c}^p=f_{B_c}^{a,0}>f_{B_c^*}^{v,i}=f_{B_c^*}^{t,i0}>f_{B_c^*}^{t5,ij}>f_{B_{c0}^*}^{v,0}>f_{B_{c0}^*}^{s}>f_{B_{c1}}^{a,i}\gtrsim f_{B_{c1}}^{t5,i0}\gtrsim f_{B_{c1}}^{t,ij}.
\end{align}

\end{enumerate}

%prediction for the ratios between  decay constants of $c\bar b$ mesons involving various currents
%the decay constants of the scalar, pseudo-scalar, vector and axial-vector $c\bar b$ mesons coupled with ten heavy flavor-changing currents.
%which implies there  must be rather large and  unignorable difference between the wave function at the origin of ... and that of ...
%the ratio of the $(s)$ to $(v,0)$ current decay constants for the  $c\bar b$ mesons

\section{Summary~\label{Summary}}

%Based on our earlier research in ref.~\cite{Tao:2023vvf},
In this work, 
we study the ten decay constants for the $S$-wave and $P$-wave $c\bar b$ mesons ($B_c,B_c^*,B_{c0}^*,B_{c1}$) coupled with the ten heavy flavor-changing currents in eq.~\eqref{j10} involving scalar, pseudo-scalar, vector, axial-vector, tensor, axial-tensor up to N$^3$LO of $\alpha_s$ within the NRQCD factorization framework.

Building upon our earlier works~\cite{Tao:2022hos,Tao:2023mtw,Tao:2023vvf},
we complete the three-loop matching between QCD and NRQCD for the heavy flavor-changing temporal vector $(v,0)$, spatial-spatial tensor $(t,ij)$ and spatial-temporal axial-tensor $(t5,i0)$ currents, which can couple to (can produce/annihilate) the $P$-wave scalar meson $B_{c0}^*$ and axial-vector meson $B_{c1}$.
We obtain the three-loop analytic expressions  of the NRQCD current renormalization constants and corresponding anomalous dimensions    for  $(v,0)$, $(t,ij)$ and $(t5,i0)$.   
Our three-loop numerical results for the matching coefficients of currents $(v,0)$, $(t,ij)$ and $(t5,i0)$ %are found  
turn out to be nonconvergent and exhibit strong scale-dependence, which is similar to what is observed  for the other seven currents  in our  previous
 research~\cite{Tao:2022hos,Tao:2023mtw,Tao:2023vvf}. 

%Furthermore, 

With the matching coefficients for all ten currents in hand,
we obtain the N$^3$LO perturbative QCD results for the six ratios of $c\bar b$ decay constants involving	various currents by approximating them to the corresponding ratios of  matching coefficients.	
We find that the N$^3$LO QCD corrections to the six ratios of decay constants exhibit good convergence behavior and have weak scale-dependence.		
If only the $S$-wave  $c\bar b$ decay constant $f_{B_c}^p$ and $P$-wave $f_{B_{c0}^*}^{s}$ are measured experimentally, our N$^3$LO results of the six decay constant ratios can be used for predicting the values of the remaining eight decay constants.				
Furthermore, considering
 $f_{S\text{-wave}}>f_{P\text{-wave}}$, 
we provide the prediction for the hierarchical relationship among the ten decay constants of the $c\bar b$ mesons coupled with the ten heavy flavor-changing currents.
These theoretical predictions will serve as a guide and be tested in flavor physics experiments, while also contributing to the determination of fundamental parameters within the SM and the study of new physics  beyond it.

%Our results for $B_c^*$ and $B_{c1}$ decay constants involving tensor and axial-tensor currents are also of interest in the study of new physics beyond the Standard Model of Particle Physics.

\hspace{2cm}

\noindent {\bf Acknowledgments:} 
We thank   A. Onishchenko and Y. B. Yang  for many helpful discussions. 

\hspace{2cm}

%\begin{appendix}
%
%\section*{Appendix ???}
%
%\end{appendix}

\bibliographystyle{JHEP}
\bibliography{refs}

\end{document}